\documentclass[preprint,preprintnumbers,amsmath,amssymb]{revtex4}
\usepackage{amsfonts}    
\usepackage{amsmath,amssymb}
\usepackage{latexsym}
\usepackage{longtable}

\newcommand{\al}{\alpha}
\newcommand{\pa}{\partial}
\newcommand{\ep}{\epsilon}

\newcommand{\ta}{\tau}

\newcommand{\Om}{\Omega}
\newcommand{\om}{\omega}
\newcommand{\de}{\delta}

\newcommand{\De}{\Delta}

\newcommand{\lrar}{\leftrightarrow}

\newcommand{\non}{\nonumber}

\begin{document}

\title{Sutherland-type Trigonometric Models, Trigonometric Invariants and
Multivariate Polynomials. III. $E_8$ case}

\author{K.G. Boreskov}
\email{boreskov@itep.ru}
\thanks{Supported in part by grants CRDF RUP2-2961-MO-09 and RFBR 08-02-00677.}
\affiliation{Institute for Theoretical and Experimental Physics, Moscow 11259, Russia}

\author{A.~V.~Turbiner}
\email{turbiner@nucleares.unam.mx}
\thanks{Supported in part by DGAPA grant IN115709-3 (Mexico)
 and the University Program FENOMEC (UNAM, Mexico)}
\affiliation{Instituto de Ciencias Nucleares, Universidad Nacional
Aut\'onoma de M\'exico, Apartado Postal 70-543, 04510 M\'exico,
D.F., Mexico}

\author{J.~C.~L\'opez Vieyra}
\email{vieyra@nucleares.unam.mx}
\thanks{Supported in part by DGAPA grant IN115709-3 (Mexico)}
\affiliation{Instituto de Ciencias Nucleares, Universidad Nacional
Aut\'onoma de M\'exico, Apartado Postal 70-543, 04510 M\'exico,
D.F., Mexico}

\author{M.A.G.~Garc\'ia}
\email{alejandro.garcia@nucleares.unam.mx}
\thanks{Supported in part by DGAPA grant IN115709-3 (Mexico)}
\affiliation{Instituto de Ciencias Nucleares, Universidad Nacional
Aut\'onoma de M\'exico, Apartado Postal 70-543, 04510 M\'exico,
D.F., Mexico}
%


\date{December 6, 2010}

\begin{abstract}
It is shown that the $E_8$ trigonometric Olshanetsky-Perelomov Hamiltonian,
when written in terms of the Fundamental Trigonometric Invariants (FTI), is in algebraic form, i.e., has polynomial coefficients, and preserves two infinite flags of polynomial spaces marked by the Weyl (co)-vector and $E_8$ highest root (both in the basis of simple roots) as characteristic vectors. The explicit form of the Hamiltonian in new variables has been obtained both by direct calculation and by means of the orbit function technique.
It is shown a triangularity of the Hamiltonian in the bases of orbit functions
and of algebraic monomials ordered through Weyl heights.
Examples of first eigenfunctions are presented.
\end{abstract}

\maketitle

\section{Introduction}

As mentioned in \cite{BLT-trigonometric} (thereafter addressed as I), about 30 years ago, Olshanetsky and Perelomov discovered a remarkable family of quantum mechanical Hamiltonians with trigonometric potentials, which are associated to the (crystallographic) root spaces of the classical and exceptional Lie algebras (for a review, see \cite{Olshanetsky:1983}). Explicitly, the Hamiltonian associated to a Lie algebra $g$ of rank $N$, with root space $\Delta$, is
\begin{equation}
\label{H}
 {H}_\Delta = \frac{1}{2}\sum_{k=1}^{N}
 \left[-\frac{\pa^{2}}{\pa y_{k}^{2}}\right]\ + \frac{\beta^2}{8}\sum_{\al\in R_{+}}
 g_{|\al|}\frac{|\,\al|^{\,2}} {\sin^2 \frac{\beta}{2} (\al , y)}\ ,
\end{equation}
where $R_+$ is the set of positive roots of $\De$, $\beta\in \mathbb{R}$ is a parameter introduced for convenience, $g_{|\al|}=\mu_{|\al|}(\mu_{|\al|}-1)$ are coupling
constants depending only on the root length, and $y = (y_1, y_2,\ldots,y_N)$
is the coordinate vector. The configuration space here is the Weyl alcove of the root space (see \cite{Olshanetsky:1983}). The Hamiltonian (\ref{H}) has the property of complete integrability: there exists $(N-1)$ integrals of motion spanning the Abelian
algebra.

The ground state eigenfunction of (\ref{H}) and its eigenvalue are
\begin{equation}
\label{Psi_0}
  \Psi_0 (y) \ =\ \prod_{\al\in R_+}
  \left|\sin \frac{\beta}{2} (\al , y)\right|^{\mu_{|\al|}}\ ,\quad
  E_0\ =\ \frac{\beta^2}{8} {\tilde \varrho}^2  \ ,
\end{equation}
where $\tilde \varrho = \sum_{\al\in R_{+}} \mu_{|\al|} {\al}$ is the so-called `deformed Weyl vector' (see \cite{Olshanetsky:1983}, Eqs.(5.5), (6.7)).
It is known that any eigenfunction $\Psi$ has the form of (\ref{Psi_0}) multiplied by a polynomial in   exponential (trigonometric) coordinates, i.e. $\Psi = \varphi \Psi_0$ (the factorization property, see \cite{Olshanetsky:1983}). The spectrum of (\ref{H}) is a second-degree polynomial in the quantum numbers \cite{Olshanetsky:1983}. The function (\ref{Psi_0}) is the (lowest) eigenfunction of any integral as well.

\smallskip

In I it was made three definitions:
(i) A linear differential operator is algebraic if its coefficients
are polynomials,
(ii) Take a linear space of multivariate polynomials
\[
 P^{(d)}_{n, \{\vec f \}} \ = \ \langle x_1^{n_1}  x_2^{n_2} \ldots x_d^{n_d} |
           0 \leq f_1 n_1 + f_2 n_2 +\ldots + f_d n_d \leq \tilde n \rangle\ \ ,
\]
where the ${f}$'s are positive integers and $\tilde n \in \mathbb{\tilde N}$, where
$\mathbb{\tilde N} \subset \mathbb{N}$ is a set of natural numbers ordered by increase such that the Diophantine equation $f_1 n_1 + f_2 n_2 +\ldots + f_d n_d = \tilde n$ has a solution. Smallest
$\tilde n$ is always equal to zero. These numbers $\tilde n(n)$ are enumerated by integer numbers $n=0,1,2,\ldots$. In many occasions $\mathbb{\tilde N} = \mathbb{N}$.
The {\it characteristic vector} is the \hbox{$d$-dimensional} vector with components $f_i$:
\begin{equation}
\label{flag}
 \vec f= (f_1, f_2, \ldots f_d)\ ,
\end{equation}
Hence, any monomial $x_1^{n_1} x_2^{n_2} \ldots x_d^{n_d}$ corresponds to a point in $d$-dimensional lattice space $\mathbf{n} = [n_1, n_2, \ldots, n_d]$, where $n_i, i=1,2,\ldots d$ are its coordinates. To each monomial (or, equivalently, to a corresponding point in the lattice space) one can assign a {\it grading},
\begin{equation}
\label{height_f}
 \mathrm{d}_{\vec f}(\mathbf{n}) \equiv (\vec f, \mathbf{n}) = f_1 n_1 + f_2 n_2
 + \ldots + f_d n_d\ .
\end{equation}
In this definition the product of two monomials $\mathbf{n}$ and $\mathbf{m}$ gives the monomial $\mathbf{n+m}$ with the grading $\mathrm{d}_{\vec f}(\mathbf{n+m})=\mathrm{d}_{\vec f}(\mathbf{n})+\mathrm{d}_{\vec f}(\mathbf{m})$,
\noindent
and
(iii) If the infinite set of spaces $P_n\equiv {P}^{(d)}_{\tilde n, \{\vec f \}}$, $\tilde n \in  \mathbb{\tilde N}$, defined as above, can be ordered by inclusion:
\[
{P }_0 \subset  { P}_1 \subset {P}_2 \subset \ldots
 \subset  {P}_n  \subset \ldots \ ,
\]
such an object is called an {\em infinite flag (or filtration)},
and is denoted ${P}^{(d)}_{\{\vec f \}}$. If for any $n$, the codimension $(\equiv \dim {P}_{n+1}-\dim {P}_{n})$ is non-zero and minimal, such a flag is called {\it dense}.
If a linear differential operator preserves such an infinite flag, it is said to
be {\it exactly-solvable}.
It is evident that every such operator is algebraic, although the converse is not true (see \cite{Turbiner:1994}).
If the spaces $P_n$ can be viewed as the finite-dimensional representation spaces of some Lie algebra $h$, then $h$ is called the {\em hidden algebra} of the exactly-solvable operator.

Any crystallographic root space $\De$ is characterized by its fundamental weights $w_a, a=1,2,\ldots N$, where $N={\rm{rank}}(\De)$. They form a (non-orthogonal) basis $\{ w \}$, which is alternative to the standard, non-orthogonal basis of simple roots $\{ \al \}$. These two bases are dual to each other in the following sense:
\renewcommand{\theequation}{5-{\arabic{equation}}}
\setcounter{equation}{0}
\begin{gather}
\label{kronecker}
 \frac{2(\al_b ,w_a)} {(\al_b,\al_b)}\ =\
 \de_{ba},\qquad b,a\in\{1,\ldots,N\}\ .
\end{gather}
If we introduce co-root for any root $\al$ as
\[
   \al^{\vee}=\frac{2\al} {(\al,\al)}\ ,
\]
the formula (\ref{kronecker}) takes a form
\begin{equation}
\label{coroot}
  (\al^{\vee}_b ,w_a)\ =\ \de_{ba},\qquad b,a\in\{1,\ldots,N\}\ .
\end{equation}
\renewcommand{\theequation}{{\arabic{equation}}}
\setcounter{equation}{5}

One can take a fundamental weight $w_a$ and generate its orbit $\Om_a$ with elements $\omega\in\Om_a$ and orbit size $d_a$, by acting on it by all elements of the Weyl group of $\De$. By averaging the exponential function over this orbit, i.e. by computing
\begin{equation}
\label{Trig_Inv}
 \ta_{a}(y) = \sum_{\om \in\Om_a} e^{i \beta (\om , y)}\ ,
\end{equation}
one obtains a trigonometric Weyl invariant for any specified $\beta\in \mathbb{R}$.
For a given root space $\Delta$ and a fixed $\beta$, there thus exist $N$ independent trigonometric Weyl invariants $\tau_a$ generated by $N$ fundamental weights $w_a$. We call them {\it Fundamental Trigonometric Invariants} (FTI) \cite{BLT-trigonometric}.

It was shown in I that for the root spaces $A_N, BC_N, B_N, C_N, D_N, G_2, F_4, E_6$
and as well as in \cite{VGT} for $E_7$ that
 (i) a similarity-transformed (\ref{H}), namely $h_{\De} \propto \Psi_0^{-1}
(H_{\De} - E_0) \Psi_0$, acting on the space of trigonometric invariants
(i.e., the space of trigonometric orbits) is an algebraic operator,
(ii) $h_{\De}$ has infinitely-many finite-dimensional invariant subspaces which form an infinite flag of spaces of polynomials, with a certain characteristic vector, and
(iii) the eigenfunctions of the Hamiltonian (\ref{H}) are polynomials in the FTI.
The goal of this paper is to show that the same three properties hold for the $E_8$ root space. Although similar results might be seen as obtainable for $E_{8}$, a complete analysis
of this root space is absent, mainly due to technical complications
which some of the present authors thought for a long time are impossible to overpass.

\medskip

\section{The case $\De = E_8$ (generalities)}

The Hamiltonian of the trigonometric $E_8$ model is built using the
root system of the $E_8$ algebra (see (\ref{H})). This Hamiltonian in
$8$-dimensional Euclidian space with Cartesian coordinates $x_1,x_2,\ldots x_8$
looks as follows
\begin{equation}
\label{H_E8}
  {H}_{E_8} = -\frac{1}{2} \Delta^{(8)} + \frac{g \beta^2}{4} \sum_{j<i =1}^{8}
  \left[\frac{1}{\sin^{2} {\frac{\beta}{2}(x_i + x_j)}} + \frac{1}{\sin^{2}
  {\frac{\beta}{2} (x_i - x_j)}} \right]
\end{equation}
\[
+ \frac{g \beta^2}{4} \sum_{\{\nu_j\}} \frac{1}{\left[\sin^{2}
\dfrac{\beta}{4}\left({ x_8  + \sum_{j=1}^7
(-1)^{\nu_j}x_j}\right)\right]} ~, \
\]
the second summation being one over septuples $\{\nu_j\}$ where each $\nu_j = 0, 1,$ and $\sum_{j=1}^{7} \nu_j \text{~is even}$. Here $g=\nu(\nu-1)>-1/4$ is the coupling constant and $\beta$ is parameter. The configuration space is the principal $E_8$ Weyl alcove.

The ground state eigenfunction and its eigenvalue are
\begin{align}
\label{Psi_E8}
 \Psi_0 &\equiv \exp[-\Phi_0(x)] = \prod_{j<i =1}^{8}
 \sin^{\nu} \frac{\beta}{2} (x_i + x_j)\sin^{\nu} \frac{\beta}{2} (x_i - x_j)
 \prod_{\{\nu_j\}}
 \sin^{\nu} \frac{\beta}{4}\left(x_8 + \sum_{j=1}^{7} (-1)^{\nu_j}\, x_{j}\right) , \\
\ E_0\ &=\ \frac{\beta^2}{2}{\boldsymbol\varrho}^{2}\nu^2\ =\ 310 \beta^2 \,\nu^2\ ,
\end{align}
(cf.(\ref{Psi_0})), where the second product is taken over for the same set $\{\nu_j\}$ as in \eqref{H_E8} and the vector ${\boldsymbol\varrho}$ is the $E_8$ Weyl vector,
\begin{align}
\label{rho}
 {\boldsymbol\varrho}\ =\ \frac{1}{2}\sum_{\al\in R_{+}} {\al} = e_2+2 e_3+3 e_4+4 e_5+5 e_6+6 e_7+23 e_8 ~, \quad {\boldsymbol\varrho}^{2}\ =\ 620\ ,
\end{align}
where $e_i,\ i=1,2,\ldots 8$ form the orthonormal basis in the $E_8$ root space.

The main object of our study is the gauge-rotated Hamiltonian (\ref{H_E8}), with the ground state eigenfunction (\ref{Psi_E8}) taken as a gauge factor, i.e.
\begin{equation}
\label{h_E8}
 h_{\rm E_8} \ =\ -\frac{2}{\beta^2}(\Psi_{0})^{-1}({H}_{\rm E_8}-E_0)\Psi_{0} \ ,
\end{equation}
where $E_0$ is given by (9). Explicitly,
\begin{align}
\label{h_E8_x}
    h_{E_8}\ = \ \sum_{k=1}^{8} \frac{\pa^2}{\pa ({\beta x_{k}})^2}
    - 2\left(\frac{\pa\Phi_0}{\pa ({\beta x_{k}})}\right)\frac{\partial}{\pa (\beta x_{k})} ~.
\end{align}
The spectral problem is
\[
   h_{E_8}\ \varphi\ =\ \ep\ \varphi\ ,
\]
where $-\ep= {2(E-E_0)}/{\beta^2}$.

The $E_8$ root space is characterized by 8 fundamental weights, which
generate orbits of lengths ranging from 240  to 483840. We follow the ordering
of the fundamental trigonometric invariants $\tau_a$ defined by (\ref{Trig_Inv})
based on the length of the orbit $\Om_a$. This ordering is different from the ordering
based on the weights $W_a$ employed by Bourbaki \cite{Bourbaki:2002} which was also used
in the MAPLE packages COXETER and WEYL by J. Stembridge \cite{numbering}) (see Table I).

\renewcommand{\arraystretch}{1.5}
\begin{center}
\begin{table}[h]
\caption{Orbit variables FTI, fundamental weights and simple roots for $E_8$.
The weights $w_a$ are ordered in their lengths or, equivalently, in the lengths of
their orbits, the weights $W_a$ follow the Bourbaki numbering. Orbit size $d_a$
indicated.}
\begin{tabular}{|c|l|r|c|}
\hline
\mbox{orbit variable}    &  \mbox{weight vectors/simple roots } & $w_a^2$\,    &~   $d_a$ \\
\hline
$\tau_1$    &   $w_1=W_8=e_7+e_8$                                               & 2\,  &~ 240~~
\\
            &   $\qquad\qquad \alpha_{1}=e_7- e_6 $                                               &      &
\\
$\tau_2$    &   $w_2=W_1=2 e_8$                                                 & 4\,  &~ 2160~~
\\
            &   $\qquad\qquad \alpha_{2}=\frac{1}{2}(e_1-e_2-e_3-e_4-e_5-e_6-e_7+e_8)$   &      &
\\
$\tau_3$    &   $w_3=W_7=e_6+e_7+2 e_8$                                         & 6\,  &~ 6720~~
\\
            &   $\qquad \qquad\alpha_{3}=e_6- e_5 $                                               &      &
\\
$\tau_4$    &   $w_4=W_2=\frac{1}{2} (e_1+e_2+e_3+e_4+e_5+e_6+e_7+5e_8)$        & 8\,  &~ 17280~~
\\
            &   $\qquad\qquad \alpha_{4}=e_1+ e_2 $                                               &      &
\\
$\tau_5$    &   $w_5=W_6=e_5+e_6+e_7+3 e_8$                                     & 12\, &~ 60480~~
\\
            &   $\qquad\qquad \alpha_{5}=e_5- e_4 $                                               &       &
\\
$\tau_6$    &   $w_6=W_3=\frac{1}{2}(-e_1+ e_2+ e_3+ e_4+ e_5+ e_6+ e_7+7e_8)$  & 14\, &~ 69120~~
\\
            &   $\qquad \qquad\alpha_{6}=e_2- e_1 $                                                 &      &
\\
$\tau_7$    &   $w_7=W_5=e_4+e_5+e_6+e_7+4 e_8$                                 & 20\, &~ 241920~~
\\
            &   $\qquad\qquad \alpha_{7}=e_4- e_3 $                                            &      &
\\
$\tau_8$    &   $w_8=W_4=e_3+e_4+e_5+e_6+e_7+5 e_8$                             & 30\, &~ 483840~~
\\
            &   $\qquad\qquad \alpha_{8}=e_3- e_2 $    &      & \\
\hline
\end{tabular}
\end{table}
\end{center}

\clearpage

Let us note that for $\beta\in \mathbb{R}$, all $\tau$'s invariants (5) are real, ${\rm Im} \tau_a =0$ (see \cite{orb_fun}). It implies some identities
\begin{equation}
\label{identities}
 {\rm Im} \ta_{a}(y) = \sum_{\om \in \Om_a} \sin \beta (\om , y)\ =\ 0\ .
\end{equation}
It is a general feature of invariants obtained in averaging over orbits generated by
the Weyl group $W(E_8)$. The reality property holds for the algebras $A_N,B_N,C_N,D_{2K},E_7,E_8,F_4,G_2$ \cite{orb_fun} (see also I and \cite{VGT}).

Our goal is to find the explicit form of the gauge-transformed Hamiltonian (\ref{h_E8}),
$h_{E_8}(\tau)$, in terms of the FTI variables $\tau=\{\tau_a\}$,
\begin{align}
\label{h_E8_tau}
    h_{E_8}(\tau) = \sum_{a,b=1}^8 A_{ab}(\tau)
    \frac{\pa}{\pa \tau_a}\frac{\pa}{\pa \tau_b}
    + \sum_{a=1}^8 B_{a}(\tau,\nu) \frac{\pa}{\pa \tau_a}~,
\end{align}
where
\begin{align}
\label{A}
    A_{ab}(\tau) &= A_{ba}(\tau) = \sum_{k=1}^8
    \frac{\pa \tau_a}{\pa ({\beta x_{k}})}\frac{\pa \tau_b}{\pa ({\beta x_{k}})} ~, \\
    B_a(\tau, \nu) &= b_a(\tau) -2\nu c_a(\tau) ~,
\end{align}
with
\begin{align}
\label{b}
    b_a(\tau) &\ =\ \sum_{k=1}^8 \frac{\pa^2 \tau_a}{\pa ({\beta x_{k}})^2} ~, \\
\label{c}
    \nu c_a(\tau) &\ =\
    \sum_{k=1}^8 \frac{\pa \Phi_0 }{\pa ({\beta x_{k}})}\frac{\pa \tau_a}{\pa({\beta x_{k}})} ~\ .
\end{align}
We will show the operator $h_{E_8}(\tau)$ takes an algebraic form, hence
$A_{ab}(\tau)$ and $B_a(\tau, \nu)$ are polynomials in $\tau$.
They do not depend on $\beta$ due to dimensional arguments. Thus, the eigenvalues do not depend on $\beta$ as well. So, without a loss of generality one can place $\beta=1$.

Therefore we have to solve a technical problem of a change of variables from Cartesian coordinates $x$ to FTI coordinates $\tau$. We approach to this problem from two different directions:
(i) making a straightforward change of variables and
(ii) using property of decomposition of orbit products into a sum of orbits,
which we call {\it the orbit method} (see below).
Each of both approaches leads to some serious technicalities. However, since they complement each other it helps overcome difficulties. Every coefficient function was either calculated in both approaches independently and then compared, or if found in one approach it was crosschecked in another one
\footnote{
For direct method of calculation an essential simplification occurred due to the fact that all $b$-coefficients (\ref{b}) and some "first" $A$-coefficients can be sufficiently easy found in an explicit form. It provided enough information to find unambiguously a flag and fix its characteristic vector. This knowledge led immediately to a constraint on the functional form of the remaining $A$-coefficients as well as $c$-coefficients.
}.

\section{Orbit method}

In this Section we present the orbit method. The method is of a general
nature and can be developed for any semi-simple Lie algebra. We limit our presentation to the case of $E_8$ which is a main object of the paper.

\subsection*{Orbits. Notation}
\indent

Let $W=\{w_a, a=1,\dots, 8\}$ be the set of fundamental weights ordered according to their length size, $\{\Om_a\}$ be the set of the Weyl group orbits generated by the vectors $w_a$ (we call them the {\it fundamental orbits}), and $\{\tau_a\}$ be the set of new, Weyl-invariant variables:
\begin{align*}
\label{tau}
 \tau_a = \sum_{\om\in \Om_a} e^{i(\om , x)} \ ,
\end{align*}
where it coincides to (4) placing $\beta=1$. Evidently, if all $x_i=0$, the variables get equal to the length of the orbit $\tau_a=d_a$. The weights and some of their characteristics are given in Table I. Note that the orbit $\Om_1$ generated by $w_1$ contains all roots of the $E_8$ algebra -- both positive and negative ones.
There is an important property of an orbit $\Om_a$ following from the Weyl invariance
\begin{equation}
\label{w}
    \sum_{\om\in \Om_a} \om \ =\ 0 \ .
\end{equation}
This property remains valid for any orbit.

The \emph{weight lattice} $L(W)$ is defined as the $\mathbb{Z}$-span of the set
of fundamental weights $W$, and the cone of \emph{dominant weights} $L_{+}(W)$ consists of
the weights with nonnegative integer coordinates in the basis of the fundamental weights:
\begin{align}
\label{basis}
 L_+ = \bigoplus_a \mathbb{Z}^{\geq0} w_a
 =\left\{\mathbf{n} = \sum_{a=1}^8 n_a w_a = [n_1, n_2 \dots n_8] ~|\ n_a\geq 0\right\}
 ~.
\end{align}
The vector $\mathbf{n}$ written in the basis of simple roots has a form $\mathbf{n}= \sum_{i=1\dots8} a_i \alpha_i$ where $a_i$ are the so-called \emph{root coordinates} of $\mathbf{n}$. The sum $\mathrm{ht} (\mathbf{n}) \equiv \sum_i a_i$ is called
the \emph{height} of $\mathbf{n}$ \cite{Humphreys:1990}. For any root system,
in particular, for $E_8$ the height can be written as
\begin{align}
\label{ht}
  \mathrm{ht}(\mathbf{n}) = ({\boldsymbol \varrho}^{\vee},\mathbf{n})\ =\
  \sum_i c_i(\boldsymbol \varrho^{\vee}) n_i  ~,
\end{align}
where $c_i({\boldsymbol \varrho}^{\vee})$ are the root coordinates of the Weyl co-vector
$\boldsymbol \varrho$ written in a basis of simple co-roots.
For simply-laced root system any root coincides to its co-root. In this case the
height is given by
\[
  \mathrm{ht}(\mathbf{n}) = ({\boldsymbol \varrho},\mathbf{n})\ =\
  \sum_i c_i(\boldsymbol \varrho) n_i ~.
\]
Explicitly, in the $E_8$ case,
\begin{align}
\label{ht_E8}
  \mathrm{ht}(\mathbf{n}) = \sum_i c_i(\boldsymbol \varrho)n_i = 29n_1+46n_2+57n_3+68n_4+84n_5+91n_6+110n_7+135n_8 ~,
\end{align}
(cf.(\ref{height_f})), where $c_i(\boldsymbol \varrho)$ are the root coordinates of the $E_8$ Weyl vector $\boldsymbol \varrho$ in the specially ordered basis of simple roots. Such an ordering (see Table I) implies that the components of the Weyl vector are increasing. This ordering agrees with the ordering of the fundamental weights by their lengths and corresponds to the numbering of the $\tau$-variables (see Table I). Hereafter, we define the \emph{height} following (\ref{ht}) for any crystallographic root space, call it the {\it Weyl height}, $\mathrm{ht}_{\boldsymbol \varrho}(\mathbf{n})$, and omit sub-index wherever it can not cause a confusion. The height and $\boldsymbol \varrho$-grading for $\tau$-polynomials
(see Eq.(4)) do coincide for the $E_8$ Weyl vector
\[
     \mathrm{ht}(\mathbf{n})\ =\ d_{\boldsymbol \varrho}(\mathbf{n})\ .
\]

Let us introduce the following ordering of dominant weights by their Weyl heights:
\begin{align}
\label{ord_n}
 \mathbf{m}<\mathbf{n}\quad \text{if}\quad \mathrm{ht}(\mathbf{m})< \mathrm{ht}(\mathbf{n}) ~.
\end{align}
It is worth noting that the equality of heights does not imply a coincidence of the lattice points.

Each orbit contains a single dominant weight of maximal height by which it can be labeled: $\Om_\mathbf{n}\ni \mathbf{n}$.
We shall use some operations with orbits -- product of orbits and decomposition of orbit products (see \cite{orb_fun}). The product of two orbits $\Om$ and $\widetilde{\Om}$ is defined as the set of all elements $\om + \widetilde{\om}$, where $\om\in\Om$ and  $\widetilde{\om}\in\widetilde{\Om}$. The product of more than two orbits is defined similarly. Product of several fundamental orbits $\Omega_a, a=1,\dots,8$ in powers $p_a$
can be decomposed into union of orbits labeled with distinct dominant weights $\mathbf{k}$
which enter with multiplicities $\mu_{(\mathbf{p})}^{(\mathbf{k})}$:
\begin{align}
\label{product_O}
  \Om_1^{p_1}\otimes \Om_2^{p_2}\cdots \otimes \Om_8^{p_8}\ =\
        \Om_{\mathbf{p}} + \bigcup_{\mathbf{k}<\mathbf{p}} \mu_{(\mathbf{p})}^{(\mathbf{k})}\Omega_\mathbf{k} ~.
\end{align}
It is essential that the highest term of the decomposition $\Omega_{\mathbf{p}}$ enters with multiplicity 1 and other terms are generated with dominant weights of lower heights than $\mathbf{p}=[p_1, p_2 \dots p_8]$.

Let us choose some dominant weight $\mathbf{n} \in L_+$. Following \cite{orb_fun} we construct an orbit function
\begin{align}
\label{M}
M_{\mathbf{n}} = \sum_{\om \in \Om_{\mathbf{n}}}\ e^{i (\om , x)} \ ,
\end{align}
by averaging the exponential function over the orbit generated by $\mathbf{n}$.
It seems evident that any orbit function is Weyl-invariant.
Orbit functions corresponding to the algebras $A_1, B_n, C_n,D_{2k}, E_7, E_8, F_4, G_2$ are real \cite{orb_fun}.

Following \eqref{ord_n} one can make the ordering of the elements in the set $\mathcal{M}=\{M_{\mathbf{n}}\}$:
\begin{align}
\label{ord_M}
 M_{\mathbf{m}} \leq M_{\mathbf{n}} \quad \text{if} \quad \mathbf{m} \leq \mathbf{n} ~.
\end{align}
In correspondence with decomposition of orbit products \eqref{product_O} a product of several fundamental $M$-functions (see below) can be decomposed into the sum of $M$-functions as
\begin{align}
\label{product_M}
  M_1^{p_1}M_2^{p_2}\cdots M_8^{p_8} =
  M_{\mathbf{p}} + \sum_{\mathbf{k}<\mathbf{p}} \mu_{(\mathbf{p})}^{(\mathbf{k})}
  M_\mathbf{k} ~.
\end{align}
Each coefficient $\mu_{(\mathbf{p})}^{(\mathbf{k})}$ is equal to multiplicity of a given orbit $\Om_\mathbf{k}$ in the decomposition of $\mathbf{p}$-product of orbits $\Om_a$
(see \eqref{product_O}).

An orbit function $M_{\mathbf{n}}$ can also be spanned on the basis $\mathcal{T}$
of monomials $\tau^{\mathbf{p}} \equiv \tau_1^{p_1}\dots \tau_8^{p_8}$.
The orbit function $M_0$ generated by zero vector $\{ 0 \}$
is equal to 1 when eight functions $M_a$ (generated by the fundamental weights $w_a, a=1,2,\ldots 8$) are nothing but the FTI $\tau_a$ (see \eqref{Trig_Inv}),
\begin{align}
\label{M1}
M_0=1 ~,\quad
M_a = \tau_a ~, \quad a=1\dots 8\ .
\end{align}
Therefore, any monomial of higher order can be obtained through the decomposition of product of the orbit functions (cf. \eqref{product_M}):
\begin{align}
\label{M_p}
 \tau^{{\mathbf{p}}} 
 =\ M_{\mathbf{p}} + \sum_{\mathbf{k}<\mathbf{p}} \mu_{(\mathbf{p})}^{(\mathbf{k})} M_{\mathbf{k}}~, \quad \mathbf{p}=[p_1,\dots , p_8] \ .
\end{align}
The set of equations \eqref{M_p} is lower triangular, so it can be solved algebraically in order to express $M_{\mathbf{n}}$ through the variables $\tau_a$:
\begin{align}
\label{M2tau}
 M_{\mathbf{n}} &= \tau^{\mathbf{n}} + \sum_{\mathbf{i}<\mathbf{n}} \tilde{\mu}_{(\mathbf{n})}^{(\mathbf{i})}
 \tau^{\mathbf{i}} ~,
\end{align}
where $\tilde{\mu}_{(\mathbf{n})}^{(\mathbf{i})}$ are some coefficients. Therefore, the Weyl height of $M_{\mathbf{n}}$ coincides with Weyl grading of the monomial $\tau^{\mathbf{n}}$.
Thus, any monomial can be marked by the Weyl height or, differently, any orbit function can be marked by the Weyl grading. Hence, the monomials in $\tau^{\mathbf{n}}$ can be ordered by Weyl heights
in accordance to \eqref{ord_n},
\begin{align}
\label{ord_tau}
 \tau^{\mathbf{m}} \leq \tau^{\mathbf{n}} \quad \text{if} \quad \mathbf{m} \leq \mathbf{n} ~.
\end{align}
Thus, the ordering of $M_{\mathbf{n}}$ is defined by the monomial of the highest height in  \eqref{M2tau} and monomials of lower heights do not influence the ordering.

The calculation of the coefficient functions in the gauge-rotated
Hamiltonian (\ref{h_E8}) (see below) requires to know explicitly a number of decomposition
equations for products of the orbit functions $M_a, a=1,2,\ldots,8$ of the type (\ref{product_M}).
Some number of such decomposition equations for multiplicities of bilinear products
$M_a M_b$ was found in \cite{Patera97}, however, we need to calculate many more bilinear decompositions as well as trilinear, 4-linear and even some 5- and 6-linear products of $M_a$. It is worth noting that the most complicated bilinear product we encountered is $M_8^2$. Its expansion consists of all orbit functions with heights up to
$\mathrm{ht}(2 w_8)=270$. In total, this set of necessary decomposition equations has been calculated using the Maple package COXETER created by John R.~Stembridge. It consists of the 253 equations, but their expressions are too long and cumbersome to be presented in this paper. It is mentioning that these equations are useful for finding the eigenfunctions of (\ref{h_E8}), see Section IV.

\subsection*{Coefficients $\mathbf{A_{ab}}$}

It can be shown that the calculation of the coefficients $A_{ab}(\tau)$ (see \eqref{A}) is reduced to calculation of a double orbit sum,
\begin{align}
\label{A_}
A_{ab}(\tau)= - \sum_{\om\in\Om_a}\sum_{\widetilde{\omega}\in\Om_b}
(\omega\cdot \widetilde{\om})\ e^{i\,(\om+\widetilde{\omega})\cdot x} ~.
\end{align}
(since now on we denote the scalar product of $a,b$ as $(a \cdot b)$).
If $x_i=0$ for all $i=1,2,\ldots 8$, as a consequence of (\ref{w}) the
coefficient $A_{ab}(x=0)=0$. From another side, at $x_i=0$ the variable $\tau_a=d_a, a=1,2,\ldots,8$ (where $d_a$ are the orbit sizes) and, hence, the coefficients $A_{ab}$ vanish,
\begin{align}
\label{A_norm}
 A_{ab}(\tau_a=d_a)\ =\ 0 \ .
\end{align}

If in the sum \eqref{A_} the factor $(\om\cdot \widetilde{\om})$ is dropped it reduces
to the product of the orbit functions $M_{a} M_{b}$, which is equal to $\tau_a \tau_b$. From the other hand, it can be represented as a superposition of the orbit functions $M$ (see \eqref{M_p}):
\begin{align}
  \tau_a \tau_b \equiv M_{a} M_{b}= \sum_{\om\in\Om_a}\sum_{\widetilde{\om}\in\Om_b}
 \ e^{i\,(\om+\widetilde{\om})\cdot x}
  = \sum_{\mathbf{n}} \ \mu_{(a,b)}^{(\mathbf{n})}\ M_{\mathbf{n}}      ~,
\end{align}
where $\mu_{(a,b)}^{(\mathbf{n})}$ are orbit multiplicities. Some orbit multiplicities for product of two orbit functions generated by the fundamental
weight orbits were calculated in \cite{Patera97}.

Let us take two orbits $\Om_a$ and $\Om_b$ generated by fundamental weights $w_a$ and $w_b$ , respectively. Denote the product of two fundamental weights as $P=(w_a\cdot w_b)$. It can be shown that $P$ is always integer. In order to check it let us take the scalar product of two arbitrary elements from these orbits, $p_{\mathbf{n}}=(\om\cdot \widetilde{\om})$. It takes one of the values $P,P-1 \dots -P+1,-P$. For two orbits with fixed $\mathbf{n}$ and, hence, $(\om + \widetilde \om)^2={\mathbf{n}}^2$, it can be calculated that $p_{\mathbf{n}}=(\mathbf{n}^2-w_a^2-w_b^2)/2$. Therefore,
\begin{align}
\label{A(tau)}
  A_{ab}(\tau)&= - \sum_{\om\in\Om_a}\sum_{\widetilde{\om}\in\Om_b}
 (\om\cdot \widetilde{\om})\ e^{i\,(\om+\widetilde{\om})\cdot x}\non \\
 &= -P \sum_{\om\in\Om_a}\sum_{\widetilde{\om}\in\Om_b}
 \ e^{i\,(\omega+\widetilde{\omega})\cdot x} +
 \sum_{\om\in\Om_a}\sum_{\widetilde{\om}\in\Om_b}
 (P - \om\cdot \widetilde{\om})\ e^{i\,(\om+\widetilde{\om})\cdot x}\non \\
 &= - P \tau_a \tau_b + \sum_{\mathbf{n}} (P- p_{\mathbf{n}})\,
 \mu_{(a,b)}^{(\mathbf{n})}\ M_{\mathbf{n}}      ~,
\end{align}

As an example, let us calculate one of the simplest coefficient $A_{12}$.
The decomposition of $M_1 M_2$ reads as
\begin{align}
\label{12}
\tau_1\tau_2 \equiv M_1 M_2 = M_{1,1,0,0,0,0,0,0}+126 M_1 + 64 M_2 +27 M_3 + 8 M_4 ~.
\end{align}
Coefficients $P,p_{\mathbf{n}}$ are for this case
\[
P=p_{1,1,0,0,0,0,0,0}=2, \qquad p_1= -2, \qquad p_2=-1, \qquad p_3=0, \qquad p_4=1 ~.
\]
Combining \eqref{A(tau)}, \eqref{12} one gets as a result
\begin{align*}
A_{12}(\tau) &= -2\tau_1 \tau_2
- 0\cdot M_{1,1,0,0,0,0,0,0}+ 4\cdot 126 M_1 + 3\cdot64 M_2 +2\cdot 27 M_3 + 1\cdot 8 M_4\non
\\
&= -2\tau_1 \tau_2 + 504\tau_1+192\tau_2 + 54\tau_3+8\tau_4 ~.
\end{align*}
For more complicated cases one has to use the expressions for $M_{\mathbf{k}}(\tau)$
from the set of relations $\mathcal{M}$ described above.

The full list of the $A_{ab}$ coefficients is given in the Appendix A.

\subsection*{Coefficients $\mathbf{b_{a}}$}

For the $b$ coefficients one can immediately derive from
eqs.\,\eqref{Trig_Inv},\,\eqref{b} that
\begin{align}
\label{B_}
     b_a\ =\ - \sum_{\om\in\Om_a} (\om\cdot \om)\ e^{i\,\om\cdot x}
     \ =\ - w_a^2 \, \tau_a \ ,
\end{align}
where $w_a^2$ is square of the fundamental weight length. In explicit form, it gives
\begin{align}
& b_{1}= -  2 \tau_1 \,,\quad
 b_{2}= -  4 \tau_2 \,,\quad
 b_{3}= -  6 \tau_3 \,,\quad
 b_{4}= -  8 \tau_4 \,,\non \\
& b_{5}= - 12 \tau_5 \,,\quad
 b_{6}= - 14 \tau_6 \,,\quad
 b_{7}= - 20 \tau_7 \,,\quad
 b_{8}= - 30 \tau_8 \,.
\end{align}

We calculated the non-interacting part of the (gauge-transformed) Hamiltonian
written in FTI variables, ${h}_{E_8}(\nu=0) = \De^{(8)}(\tau)$.
In general, any orbit function $M_\mathbf{n}$ is an eigenfunction of the Laplacian $\De^{(8)}(\tau)$ \cite{orb_fun}
\begin{align}
\label{eigen_Lap}
 \De^{(8)}(\tau) M_\mathbf{n} = - (\mathbf{n} \cdot \mathbf{n}) M_\mathbf{n} ~,
\end{align}
where $\mathbf{n}=[n_1,n_2,\dots,n_8]$ is a dominant weight. The lowest polynomial eigenfunction $\varphi_0 = \mbox{const}$ has the vanishing eigenvalue, $\ep_0=0$.
Since any $\tau$ variable is an orbit function, among the eigenfunctions
there exist eight of them which are proportional to $\tau$'s
\begin{align}
\label{nu=0}
   \varphi_{a}^{(\nu=0)} = \tau_{a} ~ ,\quad
   \epsilon_{a}^{(\nu=0)} = -w^2_{a} ~ \quad
   a = 1,2,\ldots 8\ .
\end{align}
Their eigenvalues are square of the respective fundamental weight lengths.
Following a classification based on the growth of eigenvalues $|\ep|$ at $\nu=0$ these eigenfunctions in (\ref{eigen_Lap}) have numbers 2, 3, 4, 6, 8, 10, 16, 27 (in this case to the ground state it is assigned number 1).

\subsection*{Coefficients $\mathbf{c_{a}}$}
\label{c_coefs}

According to Eqs.\eqref{Psi_0} and \eqref{c} the coefficients
$c_a$ are
\begin{align}
\label{C_}
 c_a (\tau) = - \frac{i}{2} \sum_{\alpha\in R_+}\sum_{\omega\in\Omega_a}
 (\alpha\cdot\omega)\cot \frac{(\alpha\cdot x)}{2}\, e^{i(\omega\cdot x)} ~.
\end{align}
The scalar product $l=(\alpha\cdot \omega)$ takes integer values
ranging from $-L_a$ till $L_a$ where the largest value
$L_a=(w_1\cdot w_a)$ is the $a$th root coordinate of the $E_8$
highest root $(2,2,3,3,4,4,5,6)$ in the basis of ordered simple roots
(see Table I).

In order to represent the expression \eqref{C_} as a sum of orbit
functions we will follow a construction proposed in
\cite{Sasaki:2000}. Let us consider in the sum \eqref{C_} for
every $\al$ two terms with weights $\om$ and $\hat{\om}$
related with reflection $r_{\al}$:
\begin{align*}
 \hat{\om}&=r_{\al} \om\ =\ \om - l\al ~ , \quad l=(\al\cdot\om) ~,
\end{align*}
so that ($\al^2=2$)
\begin{align}
 (\al\cdot \hat{\om})\ =\ -(\al \cdot \om) \ .
\end{align}
Let us assume for definiteness that $(\al\cdot \om)=l>0$.
Then a contribution of these two terms to \eqref{C_} is of the form
\begin{align}
\label{C_aux}
 \frac{l}{2}\, \frac{1+e^{-i(\alpha\cdot x)}}{1-e^{-i(\al\cdot x)}}
 (1-e^{-i\,l(\al\cdot x)}) e^{i(\om\cdot x)}  
  = \frac{l}{2} \left(e^{i(\om\cdot x)}+e^{i(\hat{\om}\cdot x)}\right)
 + l\left(\sum_{k=1}^{l-1} e^{i((\om - k\al)\cdot x)}\right) \ .
\end{align}
At $x=0$ it gives $l^2=(\alpha\cdot \omega)^2$.

For $(\al \cdot\om)=-l < 0$ the contribution is the same as \eqref{C_aux}
with $l=|(\al \cdot\om)|$. So (41) can be written symmetrically in
$\om$ and $\hat{\om}$ and after summation over $\om$ looks as
\begin{align}
\label{C_aux2}
 \frac{1}{2}\sum_{\om\in\Om_a} \left( l\, e^{i(\om \cdot x)}
 +l\,\sum_{k=1}^{l-1} e^{i((\om - k\al)\cdot x)}\right) ~ , ~ l=|(\al \cdot \om)| ~.
\end{align}
The first term in \eqref{C_aux2} being summed over all $\al$'s
gives
\begin{align*}
 \sum_{\om \in \Om_a} e^{i(\om \cdot x)}\sum_{\alpha\in R_+}|(\al \cdot \om)|
 = \tau_a \sum_{\al \in R_+}(\al\cdot w_a) = 2(\boldsymbol \varrho\cdot w_a)\, \tau_a ~,
\end{align*}
where
\begin{align*}
 (\boldsymbol \varrho\cdot w_a) = 29, 46, 57, 68, 84, 91, 110, 135 ~, \quad a=1,2,\dots ,8~,
\end{align*}
are the root coordinates of the Weyl vector $\boldsymbol \varrho$.

The second term in \eqref{C_aux2} contains elements of orbit
functions $M_{\mathbf{n}}$ corresponding to orbits generated by
elements $\mathbf{n}$ of the length squared $(\mathbf{n}\cdot
\mathbf{n})=(\omega-k\al)^2=w_a^2-2kl+2k^2$. The orbits
$\Om_{\mathbf{n}}$ and their multiplicities $\mu_{\mathbf{n}}$
are listed in the Table of Appendix B. Because of the symmetry
between $\omega$ and $\hat{\omega}$ the $k$-th and $(l-k)$-th
terms give equal contributions and are related to the same orbit.
After summing over $\alpha$'s the second term gives
\begin{align*}
 \frac{l}{2} \sum_{\mathbf{n}} \mu_{\mathbf{n}} M_{\mathbf{n}} ~.
\end{align*}

The coefficients $c_a$ are listed below,
\begin{align}
c_1 & = 240 + 29\,\tau_1 ~,
\\ \non
c_2 & = 126\,\tau_1 + 46\,\tau_2 ~,
\\ \non
c_3 & =168\,\tau_1+84\,\tau_2+57\,\tau_3~,
\\ \non
c_4 & =192\,\tau_2+72\,\tau_3+68\,\tau_4~,
\\
c_5 &
=-7560\,\tau_1-3672\,\tau_2-1512\,\tau_3-312\,\tau_4+84\,\tau_5+60\,\tau_1\,\tau_2~,
\non \\
c_6 &=-12096\,\tau_1 - 6144\,\tau_2 - 2448\,\tau_3 - 656\,\tau_4 +
40\,\tau_5 + 91\,\tau_6 + 96\,\tau_1\,\tau_2~,
\non \\
c_7 &= -14515200 - 5231520\,\tau_1 - 1715040\,\tau_2 -
462600\,\tau_3 - 85440\,\tau_4 + 4280\,\tau_5 + 525\,\tau_6
\non \\  &
+ 110\,\tau_7
 + 60480\,\tau_1^2 + 14880\,\tau_1\,\tau_2 - 1080\,\tau_1\,\tau_3
- 175\,\tau_1\,\tau_4 + 40\,\tau_2\,\tau_3~,
\non \\
c_8 &= 1221350400 + 440847360\,\tau_1 + 147717360\,\tau_2 +
40671720\,\tau_3 + 7663040\,\tau_4 - 387480\,\tau_5
\non \\ &
 - 52435\,\tau_6 - 1985\,\tau_7 + 135\,\tau_8 - 7644672\,\tau_1^2 -
2343552\,\tau_1\,\tau_2 - 95256\,\tau_1\,\tau_3 -
8583\,\tau_1\,\tau_4
\non \\ &
+ 1952\,\tau_1\,\tau_5
 + 204\,\tau_1\,\tau_6
- 66144\,\tau_2^2 - 17664\,\tau_2\,\tau_3 - 399\,\tau_2\,\tau_4 +
24\,\tau_2\,\tau_5 - 648\,\tau_3^2
\non \\ &
 - 84\,\tau_3\,\tau_4 +
36288\,\tau_1^3 + 9024\,\tau_1^2\,\tau_2 ~.
\non
\end{align}
It is worth noting certain relations which occur at $x_i=0$ where $\tau_a=d_a$,
\begin{align}
\label{C_norm}
 c_a(\tau_a=d_a) = \frac{d_a}{2}  \sum_{\alpha\in R_+}(\alpha\cdot w_a)^2 ~.
\end{align}

It ends the calculation of the coefficient functions of the
gauge-rotated Hamiltonian (\ref{h_E8}) using the orbit method.

\section{Spectrum and wave functions}
\label{wf}

There exist two natural bases in the infinite-dimensional functional space where the Hamiltonian \eqref{h_E8_tau} acts. The first basis, $\mathcal{T}$, consists of all monomials in $\tau_a$:
\begin{align}
 \mathcal{T} = \{ \tau^{\mathbf{n}}\equiv \prod_{a=1}^8 \tau_a^{n_a} ,
 \quad \mathbf{n}=[n_1,\dots, n_8],\quad n_a\in \mathbb{N} \} ~.
\end{align}
The second basis, $\mathcal{M}$, consists of orbit functions $M_\mathbf{n}$.
In both bases the ordering of basic elements can be introduced following the $\boldsymbol \varrho$-grading and height according to Eq.\eqref{ord_M}, \eqref{ord_tau}, respectively.

In the $\mathcal{M}$-basis any function $M_{\mathbf{n}} \in \mathcal{M}$ is an eigenfunction of the 8-dimensional Laplacian \cite{Patera97} (see \eqref{eigen_Lap}).
Furthermore, the Hamiltonian is (lower) triangular if the basic elements $M_{\mathbf{n}}$ ordered by Weyl heights (\ref{ord_n}). Thus, any eigenfunction is a linear superposition
of $M_{\mathbf{m}}$,
\begin{equation}
\label{wfM}
 \varphi_{\mathbf{n}} = M_{\mathbf{n}} + \sum_{\mathbf{m}<\mathbf{n}} c_{\mathbf{m}} M_{\mathbf{m}} ~.
\end{equation}
containing single element $M_{\mathbf{n}}$ with the highest height $\mathrm{ht}(\mathbf{n})$.
Hence, any eigenstate of the $E_8$ problem can be labeled by 8-tuple $\mathbf{n}$,
\begin{align}
 \label{eigenvalues}
 h_{E_8}\varphi_{\mathbf{n}} = \ep_{\mathbf{n}} \varphi_{\mathbf{n}},
 \quad \ep_{\mathbf{n}} = - (\mathbf{n}\cdot (\mathbf{n}+2\nu \boldsymbol \varrho)) ~,
\end{align}
(for a discussion see \cite{Sasaki:2000}, Eqs. (3.42), (3.44)) for any dominant weight $\mathbf{n}$, where $\boldsymbol \varrho$ is the Weyl vector. Let us demonstrate that (\ref{wfM}) holds.

Let us write the gauge-rotated Hamiltonian as $h_{E_8} = \De^{(8)}(\tau) + h_{int}(\tau, \nu)$, where flat-space Laplacian $\De^{(8)}(\tau)$ is given by (\ref{h_E8_tau}),(\ref{A}),(\ref{b}). The Laplacian $\De^{(8)}(\tau)$ is diagonal in $\mathcal{M}$ basis with the eigenvalues given by $- (\mathbf{n} \cdot \mathbf{n})$ (see (\ref{eigen_Lap})).
The interaction part of the Hamiltonian ${h}_{int}$  is non-diagonal in the $\mathcal{M}$ basis
\begin{align}
 h_{int} M_{\mathbf{n}}\ =\ \sum_{} c_{\mathbf{m}} M_{\mathbf{m}} ~,
\end{align}
for any $M_{\mathbf{n}} \in \mathcal{M}$. Now we can show that ${h}_{int}$ is lower triangular in this basis: $\mathbf{m}\leq\mathbf{n}$. Indeed, according to Eq.\eqref{h_E8_x},
\begin{align}
 h_{int} M_{\mathbf{n}} = -2 \sum_{k=1}^8 \left( \frac{\pa \Phi_0}{\pa x_k} \right)
 \frac{\pa}{\pa x_k} M_{\mathbf{n}} ~,
\end{align}
which is similar to the expression \eqref{c} for the coefficient $c_a$ at $\beta=1$ after a replacement of $\tau_a$ by $M_{\mathbf{n}}$.
Thus, using the same arguments as ones used for calculating the coefficients $c_a$ in Section \ref{c_coefs} one can show that the action of ${h}_{int}$
on the function $M_{\mathbf{n}}$ gives the answer similar to the answer for the coefficient $c_a$. The only difference which occurs is that the sum in $\om$ in (\ref{C_}) is taken over the orbit $\Om_{\mathbf{n}}$ instead of $\Om_a$:
\begin{align}
 h_{int} M_{\mathbf{n}} &= - \nu \{2(\boldsymbol \varrho\cdot \mathbf{n}) M_{\mathbf{n}} +
   \sum_{\al\in R_+}\sum_{\om \in\Om_{\mathbf{n}}}
   l\sum_{k=1}^{l-1} e^{i((\om -k\al)\cdot x)} \}\\
 & = - \nu \{2(\boldsymbol \varrho\cdot \mathbf{n}) M_{\mathbf{n}}
  + \sum_{\mathbf{m}<\mathbf{n}} (\al\cdot \mathbf{m}) \mu_{\mathbf{m}}M_{\mathbf{m}}\}\ , \non
\end{align}
where $l=|(\al \cdot \om)|$. Here $\mu_{\mathbf{m}}$ which occurs in the last sum is the multiplicity of the orbit generated by $\mathbf{m}$. It is seen from this formula that the interaction Hamiltonian ${h}_{int}$ has a lower triangular form and its contribution to eigenvalue is given by  $- 2\nu(\boldsymbol \varrho\cdot \mathbf{n})$ (cf. \eqref{eigenvalues}). Hence, any eigenfunction marked by $\mathbf{n}$ has a form (\ref{wfM}) and can be calculated by algebraic means.

In the $\mathcal{T}$ basis the Hamiltonian has also triangular form due to the fact that
any $M_{\mathbf{n}}$ can be written as $\tau^{\mathbf{n}}$ and
a superposition of the monomials $\tau^{\mathbf{m}}$ with $\mathrm{ht}(\mathbf{m}) < \mathrm{ht}(\mathbf{n})$ (see (\ref{M2tau})). Hence, the action of the Hamiltonian $h_{int}$ on the $\tau^{\mathbf{n}}$ gives $-2\nu(\boldsymbol \varrho,{\mathbf{n}}) \tau^{\mathbf{n}}$ and a combination of basis vectors $\tau^{\mathbf{m}}$ with Weyl heights $\mathrm{ht}(\mathbf{m}) < \mathrm{ht}(\mathbf{n})$. It implies that the results of action $h_{int}\tau^{\mathbf{n}}$ gives monomials $\tau^{\mathbf{m}}$ which all but $\tau^{\mathbf{n}}$ are situated below the hyperplane with a normal
\begin{equation}
\label{flag_orbit}
 \vec{f}_{orbit}=(29,46,57,68,84,91,110,135)\ ,
\end{equation}
while $\tau^{\mathbf{n}}$ is situated on the hyperplane.
In terms of the flag definition \eqref{flag} this means that the Hamiltonian supports the flag with the characteristic vector $\vec{f}_{orbit}=(29,46,57,68,84,91,110,135)$.
This characteristic vector is nothing but the Weyl vector $\boldsymbol \varrho$ written in the basis of simple roots. Hence, the eigenfunction marked by $\mathbf{n}$ which corresponds to the $\mathbf{n}$-th state is of the form
\begin{align}
\label{wfT}
 \varphi_{\mathbf{n}} = \tau^{\mathbf{n}} + \sum_{\mathbf{m}<\mathbf{n}} c_{\mathbf{m}} \tau^{\mathbf{m}} ~,
\end{align}
and can be calculated algebraically (cf.(\ref{wfM})).
Hence, the Hamiltonian is in triangular form in the space ${P}^{(8)}_{n,\vec{f}_{orbit}}$
with Weyl ordering of the elements. It is diagonal in action on the hyperplane defined by vector
$\vec{f}_{orbit}$,
\[
 {\cal Q}^{(8)}_{n,{\vec{f}_{orbit}}} =
\]
\begin{equation}
\label{monomials_W}
  \left\{
    \tau^{\mathbf{n}}\,  \vert
    29n_1 + 46n_2 +57n_3 +68n_4 +84n_5 +91n_6 +110n_7 + 135n_8 = \tilde n\right\}\,.
\end{equation}

The flag ${P}^{(8)}_{\vec{f}_{orbit}}$ which was found in the orbit method is not unique.
By direct calculation one can find that the operator (\ref{h_E8_tau}) preserves
the infinite flag ${P}^{(8)}_{\{ 2,2,3,3,4,4,5,6 \}}$. Its characteristic
vector
\begin{equation}
\label{flag_orbitmin}
 \vec{f}_{min} = (2,2,3,3,4,4,5,6)\ ,
\end{equation}
coincides with the $E_8$ highest root written in the basis of simple roots.
None of the characteristic vectors $\vec{f}_{orbit}$ and $\vec{f}_{min}$ coincide
with the characteristic vector in the rational $E_8$ model
$\vec{f}_{rat}=(1,3,5,5,7,7,9,11)$ found in \cite{BLT-rational}.
Furthermore, none of these two flags is supported by the rational $E_8$ Hamiltonian.
Note that this coincidence was observed for {\it all} (!) other rational and trigonometric,
crystallographic Hamiltonians \cite{BLT-trigonometric}, \cite{VGT}, \cite{BLT-rational}.

It is worth mentioning that flag of polynomials with the "absolutely" minimal characteristic vectors which might exist in $N$-dimensional space is characterized by the vector
\begin{equation}
\label{flag_orbit_0}
 \vec{f}_{0} = (1,1,\ldots, 1)\ .
\end{equation}
Sometimes it is called {\it basic}. Each subspace of the flag is a
finite-dimensional representation space of the algebra $gl(N+1)$
realized by the first order differential operators. This flag is
preserved by the rational-trigonometric Hamiltonians of $A_N$ and
$BC_N$ models (see \cite{Ruhl:1995}, \cite{Brink:1997} and a
discussion in I). One can measure a "closeness" to this flag of
any other flag $\vec{f}$ by calculating the angle between
characteristic vectors. The angles between $\vec{f}_{orbit}$ and
$\vec{f}_{min}$, and of the basic one $\vec{f}_{0}$ is given by
\[
\cos\theta_{orbit}\ =\ \frac{155}{\sqrt{28246}}\quad\simeq 0.922\ ,
\]
and
\[
\cos\theta_{min}\ =\ \frac{29}{2\sqrt{238}}\quad\simeq 0.940\ ,
\]
respectively. A fact that $\theta_{min} < \theta_{orbit}$ explains why the flag
with $\vec{f}_{min}$ is minimal. Each subspace ${P}^{(8)}_{n,{\vec{f}_{min}}}$
of the minimal flag has an outstanding property - it contains exactly
${\cal N} = \dim {\cal Q}^{(8)}_{n,{\vec{f}_{min}}}$ eigenfunctions which
involves one or more monomials from the boundary hyperplane defined by $\tilde n = n$,
\begin{equation}
\label{monomials}
  {\cal Q}^{(8)}_{n,{\vec{f}_{min}}} = \left\{\tau^{\mathbf{n}}\,  \vert
  2n_1 + 2n_2 +3n_3 +3n_4 +4n_5 +4n_6 +5n_7 + 6n_8 = n\right\}\,,
\end{equation}
where $n=0,2,3,4,\ldots$. All monomials from ${\cal Q}^{(8)}_{n,{\vec{f}_{min}}}$ have the same $\vec{f}_{min}$-grading (see (\ref{height_f})), $\mathrm{d}_{\vec f_{min}}(\mathbf{n})$. From another side, any such monomial has the Weyl grading $\mathrm{d}_{\vec f_{orbit}}(\mathbf{n})$. Triangularity of the Hamiltonian (\ref{h_E8_tau}) with respect to the Weyl-ordered monomials implies that any element of ${\cal Q}^{(8)}_{n,{\vec{f}_{min}}}$ with the Weyl grading $\mathrm{d}_{\vec f_{orbit}}(\mathbf{n})$ defines an eigenfunction which contains (or may contain) a number of monomials from ${\cal Q}^{(8)}_{n,{\vec{f}_{min}}}$ with lower Weyl grading. Since this hyperplane contains
exactly  ${\cal N}=\dim {\cal Q}^{(8)}_{n,{\vec{f}_{min}}}$ monomials, it gives rise exactly to ${\cal N}$ eigenfunctions.
\noindent
Therefore, each monomial in (\ref{monomials}) represents the highest degree term
in an eigenpolynomial from ${P}^{(8)}_{n,{\vec{f}_{min}}}$ with characteristic vector  $\vec{f}_{min}=(2,2,3,3,4,4,5,6)$. It can be marked by its degrees $\{ n_1,\ldots, n_8\}$.

Acting by the Hamiltonian $h_{E_8}(\tau)$ on the monomial of the degrees $\{n_1,\ldots, n_8\}$ we get the eigenvalue by finding a coefficient in front of
the same monomial in the r.h.s. It gives by the following explicit expression for spectrum
\begin{align}
\label{spectrum}
-\ep_{\{n_1,\ldots, n_8\}}&= 2\big[
n_1^2+2n_1n_2+3n_1n_3+3n_1n_4+4n_1n_5+4n_1n_6+5n_1n_7+6n_1n_8
\nonumber \\&
+2n_2^2+4n_2n_3+5n_2n_4+6n_2n_5+7n_2n_6+8n_2n_7+10n_2n_8
\non \\&
+3n_3^2+6n_3n_4+8n_3n_5+8n_3n_6+10n_3n_7+12n_3n_8
\nonumber \\&
+4n_4^2+9n_4n_5+10n_4n_6+12n_4n_7+15n_4n_8
+6n_5^2+12n_5n_6+15n_5n_7+18n_5n_8
\non \\&
+7n_6^2+16n_6n_7+20n_6n_8+10n_7^2+24n_7n_8
+15n_8^2\big]
\nonumber\\&
+2\big[29n_1+46n_2+57n_3+68n_4+84n_5+91n_6+110n_7+135n_8\big]\nu \ ,
\end{align}
(cf.(\ref{eigenvalues})), where $\{n_1,\ldots n_8\}=0,1,\ldots$.
When the set $\{n_1,\ldots n_8\}$ is subject to the condition $2n_1+2n_2+3n_3+3n_4+4n_5+4n_6+5n_7+6n_8 = n$, we get eigenvalues of the eigenstates
from $P^{(8)}_{{n},{\vec{f}_{min}}}$. The first eigenvalues ordered following the
increasing values of $|(\mathbf{n} \cdot \mathbf{n})|$ are listed in Table~\ref{Etable}.
\begin{table}[h]
 \caption{\label{Etable}
 The first eigenvalues $\ep_{[n_1,\ldots, n_8]}= -(\mathbf{n} \cdot \mathbf{n})-2\nu (\mathbf{n} \cdot \boldsymbol \varrho)\ $ of the gauge rotated Hamiltonian $h_{E_8}$ ordered by the increasing  values of $|(\mathbf{n} \cdot \mathbf{n}) |$ and their corresponding
 $n=2n_1+2n_2+3n_3+3n_4+4n_5+4n_6+5n_7+6n_8 $.}
 \begin{center}
\begin{tabular}{|l|l|r|r |}\hline
  $\ep_{\mathbf n}$ & n & $(\mathbf n, \mathbf n)$ & $\mathrm{ht}(\mathbf n$)\\
\hline
     $\ep_{ \left[0\, 0\, 0\, 0\, 0\, 0\, 0\, 0\,\right]} = 0         $ &  n= 0   &
 0   & 0    \\[-15pt]
     $\epsilon_{ \left[1\, 0\, 0\, 0\, 0\, 0\, 0\, 0\,\right]} = -2-58\nu  $ &  n= 2   &
 2   & 29  \\[-15pt]
     $\epsilon_{ \left[0\, 1\, 0\, 0\, 0\, 0\, 0\, 0\,\right]} = -4-92\nu  $ &  n= 2   &
 4   & 46  \\[-15pt]
     $\epsilon_{ \left[0\, 0\, 1\, 0\, 0\, 0\, 0\, 0\,\right]} = -6-114\nu $ &  n= 3   &
 6   & 57  \\[-15pt]
     $\epsilon_{ \left[2\, 0\, 0\, 0\, 0\, 0\, 0\, 0\,\right]} = -8-116\nu $ &  n= 4   &
 8   & 58  \\[-15pt]
     $\epsilon_{ \left[0\, 0\, 0\, 1\, 0\, 0\, 0\, 0\,\right]} = -8-136\nu $ &  n= 3   &
 8   & 68  \\[-15pt]
     $\epsilon_{ \left[1\, 1\, 0\, 0\, 0\, 0\, 0\, 0\,\right]} = -10-150\nu$ &  n= 4   &
 10  & 75  \\[-15pt]
     $\epsilon_{ \left[0\, 0\, 0\, 0\, 1\, 0\, 0\, 0\,\right]} = -12-168\nu$ &  n= 4   &
 12  & 84  \\[-15pt]
     $\epsilon_{ \left[1\, 0\, 1\, 0\, 0\, 0\, 0\, 0\,\right]} = -14-172\nu$ &  n= 5   &
 14  & 86  \\[-15pt]
     $\epsilon_{ \left[0\, 0\, 0\, 0\, 0\, 1\, 0\, 0\,\right]} = -14-182\nu$ &  n= 4   &
 14  & 91  \\[-15pt]
     $\epsilon_{ \left[0\, 2\, 0\, 0\, 0\, 0\, 0\, 0\,\right]} = -16-184\nu$ &  n= 4   &
 16  & 92  \\[-15pt]
     $\epsilon_{ \left[1\, 0\, 0\, 1\, 0\, 0\, 0\, 0\,\right]} = -16-194\nu$ &  n= 5   &
 16  & 97  \\[-15pt]
     $\epsilon_{ \left[0\, 1\, 1\, 0\, 0\, 0\, 0\, 0\,\right]} = -18-206\nu$ &  n= 5   &
 18  & 103 \\[-15pt]
     $\epsilon_{ \left[3\, 0\, 0\, 0\, 0\, 0\, 0\, 0\,\right]} = -18-174\nu$ &  n= 6   &
 18  & 87  \\[-15pt]
     $\epsilon_{ \left[2\, 1\, 0\, 0\, 0\, 0\, 0\, 0\,\right]} = -20-208\nu$ &  n= 6   &
 20  & 104 \\[-15pt]
     $\epsilon_{ \left[0\, 0\, 0\, 0\, 0\, 0\, 1\, 0\,\right]} = -20-220\nu$ &  n= 5   &
 20  & 110 \\[-15pt]
     $\epsilon_{ \left[1\, 0\, 0\, 0\, 1\, 0\, 0\, 0\,\right]} = -22-226\nu$ &  n= 6   &
 22  & 113 \\[-15pt]
     $\epsilon_{ \left[0\, 1\, 0\, 1\, 0\, 0\, 0\, 0\,\right]} = -22-228\nu$ &  n= 5   &
 22  & 114 \\[-15pt]
     $\epsilon_{ \left[0\, 0\, 2\, 0\, 0\, 0\, 0\, 0\,\right]} = -24-228\nu$ &  n= 6   &
 24  & 114 \\[-15pt]
     $\epsilon_{ \left[1\, 0\, 0\, 0\, 0\, 1\, 0\, 0\,\right]} = -24-240\nu$ &  n= 6   &
 24  & 120 \\[-15pt]
     $\epsilon_{ \left[2\, 0\, 1\, 0\, 0\, 0\, 0\, 0\,\right]} = -26-230\nu$ &  n= 7   &
 26  & 115 \\[-15pt]
     $\epsilon_{ \left[1\, 2\, 0\, 0\, 0\, 0\, 0\, 0\,\right]} = -26-242\nu$ &  n= 6   &
 26  & 121 \\[-15pt]
     $\epsilon_{ \left[0\, 0\, 1\, 1\, 0\, 0\, 0\, 0\,\right]} = -26-250\nu$ &  n= 6   &
 26  & 125 \\[-15pt]
     $\epsilon_{ \left[2\, 0\, 0\, 1\, 0\, 0\, 0\, 0\,\right]} = -28-252\nu$ &  n= 7   &
 28  & 126 \\[-15pt]
     $\epsilon_{ \left[0\, 1\, 0\, 0\, 1\, 0\, 0\, 0\,\right]} = -28-260\nu$ &  n= 6   &
 28  & 130 \\[-15pt]
     $\epsilon_{ \left[1\, 1\, 1\, 0\, 0\, 0\, 0\, 0\,\right]} = -30-264\nu$ &  n= 7   &
 30  & 132 \\[-15pt]
     $\epsilon_{ \left[0\, 0\, 0\, 0\, 0\, 0\, 0\, 1\,\right]} = -30-270\nu$ &  n= 6   &
 30  & 135 \\[-15pt]
     $\epsilon_{ \left[4\, 0\, 0\, 0\, 0\, 0\, 0\, 0\,\right]} = -32-232\nu$ &  n= 8   &
 32  & 116 \\[-15pt]
     $\epsilon_{ \left[3\, 1\, 0\, 0\, 0\, 0\, 0\, 0\,\right]} = -34-266\nu$ &  n= 8   &
 34  & 133 \\
 \hline
 \end{tabular}
 \end{center}
\end{table}

The $E_8$ Hamiltonian depends on the parameter $\nu$. Thus, the nodal structure of an eigenfunction (where it vanishes) depends on $\nu$ as well. In general, the nodal structure
at fixed $\nu$ as well as its evolution with $\nu$ remains an open question.

It is worth noting that there exists a degeneracy of the eigenstates for any $\nu$.
We do not know how to classify the degenerate states since it requires to solve
a system of two Diophantine equations, one linear and one quadratic, of eight variables.
The degenerate states are not present among first 29 eigenstates (see below Table II). The first degenerate states are $[0\, 2\, 1\, 0\, 0\, 0\, 0\, 0]$ and $[2\, 0\, 0\, 0\, 0\, 1\, 0\, 0]$ with the eigenvalue $\ep_{\mathbf n}= - 38 - 298\nu$. They are among eigenstates with numbers 33 and 37. Their eigenfunctions contain monomials of the highest heights $\tau_2^2 \tau_3$ and $\tau_1^2 \tau_6$, respectively.

\clearpage

\section{Conclusion}

Weyl-invariant coordinates FTI leading to the algebraic forms of the trigonometric Olshanetsky-Perelomov Hamiltonians associated to the crystallographic root spaces
$A_N, BC_N, G_2, F_4, E_6$ and $E_7$ were found in \cite{Ruhl:1995}, \cite{Brink:1997}, \cite{Rosenbaum:1998}, \cite{blt}, \cite{BLT-trigonometric}
(see I for a discussion as well as \footnote{There is a bug in MAPLE-11 (and the subsequent versions of MAPLE): the operator "sum" gives regularly wrong but stable results. Due to this fact the coefficients $B_a$ (see (13)) for the $E_6$ trigonometric model were calculated wrongly. Recently, they were recalculated in the orbit method, and crosschecked in MAPLE-13 with use of the operator "add" instead of "sum" and also in MAPLE-8
with use "sum". Correct coefficients are the following
$$
  B_1\ =\ -\frac{4}{3}(1+12\nu) \tau_1\ ,\ B_2\ =\ -\frac{4}{3}\tau_2 -2\nu (11 \tau_2 + 72)\ ,\
  B_3\ = -\frac{10}{3}\tau_3-10\nu (3 \tau_3 + 8\tau_6)\ ,
$$
$$
  B_4\ =\ -\frac{10}{3}\tau_4-6\nu (7\tau_4 + 8\tau_1 \tau_6 - 28\tau_2 - 216)\ ,\
  B_5\ = -2\tau_5 -10\nu (3\tau_5 + 8\tau_1)\ ,\
  B_6\ =\ -2(3 + 8\nu) \tau_6\ ,
$$
(cf. I, p.29). The symmetry of B-coefficients w.r.t. $\tau_1\lrar\tau_2,\tau_3\lrar\tau_4$ exists for $\nu=0$ only (cf.I, p.29)
})
and \cite{VGT}, respectively. In this paper, we have shown that the fundamental trigonometric invariants (FTI), if used as coordinates, provide a way
of reducing the trigonometric Hamiltonian associated to $E_8$ to algebraic form.
The eigenfunctions of the trigonometric Hamiltonian $E_8$ (i.e., the $E_8$ Jack polynomials) remain polynomials in the FTI. The use of FTI enabled us to find an algebraic form of the Hamiltonian associated to $E_8$, which did not seem feasible at all, in the past. The calculations in this paper were carried out in two different ways: (i) with a straightforward change of variables from Cartesian coordinates to FTI and (ii) using the orbit method. These two approaches have led to two different flags of finite-dimensional invariant subspaces of the $E_8$ Hamiltonian described by the characteristic vectors $(2,2,3,3,4,4,5,6)$ and $(29,46,57,68,84,91,110,135)$, respectively. The first vector coincides with the $E_8$ highest root in the basis of simple roots and the second one is the $E_8$ Weyl vector spanned in simple roots. In the orbit method it was easily demonstrated that the obtained algebraic Hamiltonian is in triangular form with respect to the action over a set of
 ordered monomials following the $E_8$ Weyl grading.

So far, each of the Olshanetsky-Perelomov Hamiltonians, in algebraic form, preserves an infinite flag of polynomial spaces, with a characteristic vector ${\vec f}$ that coincides with the minimal characteristic vector for the corresponding rational model (see \cite{BLT-rational}). The present study of the $E_8$ trigonometric case does not confirm this correspondence: none of these two vectors corresponds to the minimal vector of the $E_8$ rational model. Moreover the flag with any of those vectors is not preserved by the $E_8$ rational Hamiltonian.

The orbit method admits a straightforward generalization towards any Olshanetsky-Perelomov trigonometric Hamiltonian which is symmetric to the (crystallographic) Weyl group. It seems also as a straightforward result that the {\it integer} Weyl vector \footnote{Integer Weyl vector is minimal vector with integer coefficients proportional to the Weyl vector} spanned in the simple roots is the characteristic vector of the flag preserved by the algebraic Hamiltonian written in FTI (cf. Table III). Furthermore, the algebraic Hamiltonian is triangular with respect to the action on Weyl-grading-ordered monomials. It is checked immediately by the direct calculation. For non-simply laced cases the {\it integer} Weyl-co-vector \footnote{Integer Weyl co-vector is minimal vector  with integer coefficients proportional to the Weyl co-vector} is also the characteristic vector (see Table III).

\begin{table}[h]
 \caption{Minimal characteristic vectors for rational (non)crystallographic and
  trigonometric crystallographic systems. For latter case the Weyl vector and
  co-vector as possible characteristic vectors occur. Characteristic vectors for
  $H_3$, $H_4$, $I_2(k)$ are from \cite{H3, H4, I2K}, respectively.}
\begin{center}
\begin{tabular}{|c|c|c|c|c|}
\hline
             &           &           \multicolumn{3}{c|}{ }    \\[-15pt]
\hspace{10pt} Model \hspace{10pt} & Rational  &  \multicolumn{3}{c|}{    \raise
12pt\hbox{Trigonometric} }  \\ \cline{3-5}

&  & \hspace{5pt} Minimal &  \hspace{5pt}integer Weyl& \hspace{5pt} integer co-Weyl\hspace{5pt}\\[5pt]
\hline \hline

\raise 12pt\hbox{$A_N$} & $\buildrel\underbrace{(1,1,\ldots
1)}\over N$ &
$\buildrel\underbrace{(1,1,\ldots 1)}\over N$ && \\
             \hline
\raise 12pt\hbox{$BC_N$} & $\buildrel\underbrace{(1,1,\ldots
1)}\over N$ &
$\buildrel\underbrace{(1,1,\ldots 1)}\over N$ && \\
             \hline
$G_2$ & (1,2) & (1,2) &  (3,5) & $(5,9)$\\
             \hline
$F_4$ & (1,2,2,3) & (1,2,2,3) & (8,11,15,21)& (11,16,21,30)\\
             \hline
$E_6$ & (1,1,2,2,2,3) & (1,1,2,2,2,3) & (8,8,11,15,15,21)& (8,8,11,15,15,21)\\
             \hline
$E_7$ & (1,2,2,2,3,3,4) & (1,2,2,2,3,3,4) &
$(27,34,49,52,66,75,96)$&$(27,34,49,52,66,75,96)$
\\
             \hline
$E_8$ & (1,3,5,5,7,7,9,11) & (2,2,3,3,4,4,5,6) & (29,46,57,68,84,91,110,135)&
(29,46,57,68,84,91,110,135)\\
             \hline
$H_3$ & (1,2,3) & --- && \\
             \hline
$H_4$ & (1,5,8,12) & --- && \\
             \hline
$I_2(k)$ & (1,k) & --- && \\
             \hline
\end{tabular}
\end{center}
\end{table}

It is worth noting that the matrices $A_{ij}$ in the algebraic form Hamiltonians, in particular, given explicitly in Eqs. (\ref{h_E8_tau}), with polynomial entries,
correspond to flat-space metrics, in the sense that the associated Riemann tensor
vanishes. The change of variables in the corresponding Laplace-Beltrami operator,
from FTI to Cartesian coordinates, transforms these metrics to diagonal form of
the unit matrix. This procedure provides a set of non-trivial metrics with polynomial
entries with vanishing Riemann tensor.

It should be stressed that each Hamiltonian of the form (\ref{H}) is completely
integrable. This implies the existence of a number of operators (sometimes they
are called the `higher Hamiltonians') which commute with it and which are in involution forming a commutative algebra. It is evident that these commuting operators take on an algebraic form after a gauge rotation (with the corresponding ground state eigenfunction
as a gauge factor), and a change of variables from Cartesian coordinates to the FTI,
i.e., to the $\tau$'s (for a discussion, see \cite{Turbiner:1994}). An interesting open question is: are they characterized by the same flag(s) of invariant subspaces as the Hamiltonian?

In concluding, we mention that the existence of algebraic form of the $E_8$
trigonometric Olshanetsky-Perelomov Hamiltonian makes possible the study of
their perturbations by purely algebraic means. If the perturbation is of the
same Weyl symmetry as the original Hamiltonian being a polynomial in FTI,
one can develop a perturbation theory in which all corrections are found by
linear-algebraic methods \cite{Tur-pert}. It also gives a hint that quasi-exactly
solvable generalizations (see \cite{Turbiner:1988}) of the $E_8$ trigonometric Olshanetsky-Perelomov Hamiltonian may exist.

\bigskip

\textit{\small Acknowledgements}. A part of computations in this paper were performed
on MAPLE 8 while another part on MAPLE 12 with the package COXETER created by J.~Stembridge.
K.G.B. is supported in part by grants CRDF RUP2-2961-MO-09 and RFBR 08-02-00677.
One of us (A.V.T.) is grateful to IHES for its kind hospitality extended to him,
while the essential part of the present study was done. The research is supported in part
by DGAPA grant IN115709 (Mexico). A.V.T. thanks the University Program
FENOMEC (UNAM, Mexico) for partial support. J.C.L.V. thanks the
PASPA-DGAPA program (Mexico) for a partial support and the Department of Mathematics
for the Ohio State University (Columbus, Ohio) for the kind hospitality extended
to him during the sabbatical stay where a part of the present work was done.

\newpage

\section*{Appendix A}

A complete list of the $A_{ab}$ coefficients (see (\ref{h_E8}), (14) and (\ref{A})) is
\small

$ A_  {11} =  960 + 168\tau_1 + 28\tau_2 + 2\tau_3 - 2\tau_1^2\,,
$

$ A_  {12} =  504\tau_1 + 192\tau_2 + 54\tau_3 + 8\tau_4 -
2\tau_1\tau_2\,, $

$ A_  {13} =  -40320 - 12096\tau_1 - 3468\tau_2 - 768\tau_3 -
108\tau_4 + 3\tau_5 + 168\tau_1^2 + 24\tau_1\tau_2
   - 3\tau_1\tau_3\,,
$

$ A_  {14} =  -12096\tau_1 - 5760\tau_2 - 2232\tau_3 - 544\tau_4 +
32\tau_5 + 7\tau_6 + 96\tau_1\tau_2 - 3\tau_1\tau_4\,, $

$ A_ {15} =  -5806080 - 2122848\tau_1 - 701184\tau_2 -
191232\tau_3 - 35904\tau_4 + 1808\tau_5
   + 259\tau_6 + 4\tau_7 + 24192\tau_1^2 + 6168\tau_1\tau_2 -
432\tau_1\tau_3 - 77\tau_1\tau_4 -
4\tau_1\tau_5
   + 20\tau_2\tau_3\,,
$

$ A_  {16} =  12026880 + 4428864\tau_1 + 1464144\tau_2 +
401832\tau_3 + 75792\tau_4 - 3872\tau_5
   - 561\tau_6 - 15\tau_7 - 48384\tau_1^2 - 12960\tau_1\tau_2 +
1224\tau_1\tau_3 + 287\tau_1\tau_4 -
4\tau_1\tau_6
   - 192\tau_2^2 - 64\tau_2\tau_3 + 7\tau_2\tau_4\,,
$

$ A_  {17} = \ 491443200 + 175564800\tau_1 + 59465760\tau_2 +
16428960\tau_3 + 3102880\tau_4
   - 156760\tau_5 - 21135\tau_6 - 860\tau_7 + 5\tau_8 -
3786048\tau_1^2 - 1208208\tau_1\tau_2
   - 94824\tau_1\tau_3 - 12072\tau_1\tau_4 + 1248\tau_1\tau_5 +
146\tau_1\tau_6 - 5\tau_1\tau_7 - 38016\tau_2^2
   - 10096\tau_2\tau_3 - 336\tau_2\tau_4 + 16\tau_2\tau_5 -
432\tau_3^2 - 51\tau_3\tau_4 + 24192\tau_1^3
   + 5856\tau_1^2\tau_2\,,
$

$ A_  {18} =  19583078400 + 7128241920\tau_1 + 2368505760\tau_2 +
651394800\tau_3
   + 123037280\tau_4 - 6238080\tau_5 - 846165\tau_6 - 32230\tau_7
+
325\tau_8
   - 100336320\tau_1^2 - 30549744\tau_1\tau_2 + 85176\tau_1\tau_3
+
279058\tau_1\tau_4 + 5384\tau_1\tau_5
   - 1189\tau_1\tau_6 - 211\tau_1\tau_7 - 6\tau_1\tau_8 -
1191360\tau_2^2 - 505624\tau_2\tau_3 - 16106\tau_2\tau_4
   - 560\tau_2\tau_5 - 255\tau_2\tau_6 - 12\tau_2\tau_7 -
67536\tau_3^2 - 15643\tau_3\tau_4 + 432\tau_3\tau_5
   + 82\tau_3\tau_6 - 712\tau_4^2 + 30\tau_4\tau_5 + 6\tau_4\tau_6
+
399168\tau_1^3 + 83136\tau_1^2\tau_2
   + 21816\tau_1^2\tau_3 + 3955\tau_1^2\tau_4 - 5088\tau_1\tau_2^2
+
1856\tau_1\tau_2\tau_3 + 411\tau_1\tau_2\tau_4
   - 384\tau_2^3 - 96\tau_2^2\tau_3\,,
$

$ A_  {22} =  -103680 - 36288\tau_1 - 11520\tau_2 - 2880\tau_3 -
488\tau_4 + 20\tau_5 + 2\tau_6 + 504\tau_1^2
   + 96\tau_1\tau_2 -
4\tau_2^2\,, $

$ A_  {23} =  725760 + 235872\tau_1 + 72744\tau_2 + 17856\tau_3 +
2968\tau_4 - 130\tau_5 - 7\tau_6
   - 3024\tau_1^2 - 504\tau_1\tau_2 +
54\tau_1\tau_3 + 7\tau_1\tau_4 - 4\tau_2\tau_3\,, $

$ A_  {24} =  - 3317760 - 1124928\tau_1 - 354576\tau_2 -
92520\tau_3 - 16336\tau_4 + 776\tau_5
   + 57\tau_6 + 5\tau_7 + 12096\tau_1^2 + 2784\tau_1\tau_2 -
576\tau_1\tau_3 - 119\tau_1\tau_4 + 192\tau_2^2
   + 32\tau_2\tau_3 - 5\tau_2\tau_4\,,
$

$ A_  {25} =  - 3732480 - 886464\tau_1 - 487728\tau_2 -
162648\tau_3 - 34128\tau_4 + 1512\tau_5
   + 135\tau_6 + 63\tau_7 + 217728\tau_1^2 + 81936\tau_1\tau_2 +
12960\tau_1\tau_3 + 1419\tau_1\tau_4
   - 108\tau_1\tau_5 - 6\tau_1\tau_6 + 4584\tau_2^2 +
1008\tau_2\tau_3 + 21\tau_2\tau_4 - 6\tau_2\tau_5 + 54\tau_3^2
   + 6\tau_3\tau_4 - 3024\tau_1^3 - 600\tau_1^2\tau_2\,,
$

$ A_  {26} =  60134400 + 19870272\tau_1 + 7282608\tau_2 +
2094264\tau_3 + 407600\tau_4
   - 20608\tau_5 - 3036\tau_6 - 145\tau_7 + 3\tau_8 -
1112832\tau_1^2 - 419424\tau_1\tau_2
   - 61704\tau_1\tau_3 - 7715\tau_1\tau_4 + 408\tau_1\tau_5 -
6\tau_1\tau_6 - 25920\tau_2^2 - 7576\tau_2\tau_3
   -607\tau_2\tau_4 + 16\tau_2\tau_5 - 7\tau_2\tau_6 - 432\tau_3^2
-
51\tau_3\tau_4 + 12096\tau_1^3 + 2976\tau_1^2\tau_2
   + 96\tau_1\tau_2^2\,,
$
\newpage
$ A_  {27} =  - 9354009600 - 3371906880\tau_1 - 1127785200\tau_2 -
    311513400\tau_3 - 58776560\tau_4 + 2975440\tau_5 + 402480\tau_6 +
    15505\tau_7 - 135\tau_8 + 59984064\tau_1^2 + 19795152\tau_1\tau_2 +
    882864\tau_1\tau_3 + 62275\tau_1\tau_4 - 13744\tau_1\tau_5 - 
    1137\tau_1\tau_6 + 41\tau_1\tau_7 + 969600\tau_2^2 + 290248\tau_2\tau_3 +
     23483\tau_2\tau_4 - 880\tau_2\tau_5 - 35\tau_2\tau_6 - 8\tau_2\tau_7 + 
     14760\tau_3^2 + 2501\tau_3\tau_4 - 32\tau_3\tau_5 - 5\tau_3\tau_6 + 
     20\tau_4^2 + 5\tau_4\tau_5 - 314496\tau_1^3 - 87552\tau_1^2\tau_2 - 
     3456\tau_1^2\tau_3 -  623\tau_1^2\tau_4 - 2400\tau_1\tau_2^2 + 
     80\tau_1\tau_2\tau_3 + 15\tau_1\tau_2\tau_4\,,
$

$ A_  {28} =  - 234132249600 - 89159590080\tau_1 -
28698829200\tau_2 - 7858575720\tau_3
   - 1476139920\tau_4 + 74828160\tau_5 + 10111680\tau_6 +
384555\tau_7 - 4485\tau_8
   - 197752320\tau_1^2 - 245801568\tau_1\tau_2 -
157514184\tau_1\tau_3 - 29835487\tau_1\tau_4
   + 1274600\tau_1\tau_5 + 182346\tau_1\tau_6 + 8720\tau_1\tau_7 -
84\tau_1\tau_8 - 31519440\tau_2^2
   - 14864688\tau_2\tau_3 - 2730847\tau_2\tau_4 +
152584\tau_2\tau_5
+ 18034\tau_2\tau_6 + 633\tau_2\tau_7
   - 10\tau_2\tau_8 - 1476504\tau_3^2 - 392419\tau_3\tau_4 +
23488\tau_3\tau_5 + 1585\tau_3\tau_6 + 78\tau_3\tau_7
   - 19500\tau_4^2 + 2309\tau_4\tau_5 + 42\tau_4\tau_6 +
4\tau_4\tau_7 - 64\tau_5^2 - 4\tau_5\tau_6 + 27881280\tau_1^3
   + 12101472\tau_1^2\tau_2 + 1071792\tau_1^2\tau_3 +
122236\tau_1^2\tau_4 - 12096\tau_1^2\tau_5 - 936\tau_1^2\tau_6
   + 1148544\tau_1\tau_2^2 + 210608\tau_1\tau_2\tau_3 +
14563\tau_1\tau_2\tau_4 - 1040\tau_1\tau_2\tau_5 -
20\tau_1\tau_2\tau_6
   + 1296\tau_1\tau_3^2 - 18\tau_1\tau_3\tau_4 + 14\tau_1\tau_4^2
+
15168\tau_2^3 + 5248\tau_2^2\tau_3 + 35\tau_2^2\tau_4
   + 384\tau_2\tau_3^2 + 10\tau_2\tau_3\tau_4 - 217728\tau_1^4 -
67776\tau_1^3\tau_2 - 3456\tau_1^2\tau_2^2\,, $

$ A_  {33} =  - 5806080 - 2007936\tau_1 - 624480\tau_2 -
160704\tau_3 - 27968\tau_4 + 1336\tau_5
   + 84\tau_6 + 4\tau_7 + 12096\tau_1^2 + 1176\tau_1\tau_2 -
1368\tau_1\tau_3 - 178\tau_1\tau_4 + 2\tau_1\tau_5
   - 8\tau_2\tau_3 - 6\tau_3^2 + 168\tau_1^3 + 24\tau_1^2\tau_2\,,
$

$ A_  {34} =  41472000 + 14833152\tau_1 + 4813248\tau_2 +
1300608\tau_3 + 238720\tau_4
   - 11840\tau_5 - 1248\tau_6 - 68\tau_7 - 169344\tau_1^2 -
46848\tau_1\tau_2 + 3888\tau_1\tau_3
   + 632\tau_1\tau_4 + 32\tau_1\tau_5 + 6\tau_1\tau_6 -
1728\tau_2^2 - 320\tau_2\tau_3 - 6\tau_3\tau_4 +
96\tau_1^2\tau_2\,, $

$ A_  {35} =  583925760 + 204472512\tau_1 + 70943952\tau_2 +
19662552\tau_3 + 3725136\tau_4
   - 187464\tau_5 - 25306\tau_6 - 1177\tau_7 + 5\tau_8 -
6072192\tau_1^2 - 1935600\tau_1\tau_2
   - 232632\tau_1\tau_3 - 33859\tau_1\tau_4 + 2464\tau_1\tau_5 +
255\tau_1\tau_6 + 3\tau_1\tau_7 - 44880\tau_2^2
   - 10640\tau_2\tau_3 + 77\tau_2\tau_4 - 12\tau_2\tau_5 -
432\tau_3^2 - 39\tau_3\tau_4 - 8\tau_3\tau_5 + 48384\tau_1^3
   + 10752\tau_1^2\tau_2 - 432\tau_1^2\tau_3 - 77\tau_1^2\tau_4 +
20\tau_1\tau_2\tau_3\,, $

$ A_  {36} =  - 2769500160 - 975445632\tau_1 - 335008224\tau_2 -
93037680\tau_3 - 17647712\tau_4
   + 894880\tau_5 + 123760\tau_6 + 4706\tau_7 - 62\tau_8 +
26925696\tau_1^2 + 9087264\tau_1\tau_2
   + 1016640\tau_1\tau_3 + 140174\tau_1\tau_4 - 9840\tau_1\tau_5 -
798\tau_1\tau_6 - 10\tau_1\tau_7 + 370176\tau_2^2
   + 107712\tau_2\tau_3 + 7654\tau_2\tau_4 - 224\tau_2\tau_5 +
5472\tau_3^2 + 752\tau_3\tau_4 - 8\tau_3\tau_6 + 8\tau_4^2
   - 205632\tau_1^3 - 52800\tau_1^2\tau_2 + 864\tau_1^2\tau_3 +
210\tau_1^2\tau_4 - 1248\tau_1\tau_2^2 - 48\tau_1\tau_2\tau_3
   + 6\tau_1\tau_2\tau_4\,,
$

$ A_  {37} =  57683404800 + 21385226880\tau_1 + 6968748000\tau_2 +
1919657520\tau_3
   + 361844960\tau_4 - 18345280\tau_5 - 2486120\tau_6 -
95370\tau_7
+ 910\tau_8
   - 156083328\tau_1^2 - 43822944\tau_1\tau_2 +
14321664\tau_1\tau_3
+ 3234090\tau_1\tau_4
   - 106688\tau_1\tau_5 - 20162\tau_1\tau_6 - 1390\tau_1\tau_7 +
4\tau_1\tau_8 - 3926784\tau_2^2
   - 1235360\tau_2\tau_3 - 74242\tau_2\tau_4 - 576\tau_2\tau_5 -
920\tau_2\tau_6 - 40\tau_2\tau_7 - 91584\tau_3^2
   - 24476\tau_3\tau_4 + 80\tau_3\tau_5 - 38\tau_3\tau_6 -
10\tau_3\tau_7 - 1812\tau_4^2 + 45\tau_4\tau_5 + 6\tau_4\tau_6
   - 2274048\tau_1^3 - 842880\tau_1^2\tau_2 - 62064\tau_1^2\tau_3
-
4470\tau_1^2\tau_4 + 1248\tau_1^2\tau_5
   + 146\tau_1^2\tau_6 - 43104\tau_1\tau_2^2 -
9200\tau_1\tau_2\tau_3 + 370\tau_1\tau_2\tau_4 +
16\tau_1\tau_2\tau_5 - 432\tau_1\tau_3^2
   - 51\tau_1\tau_3\tau_4 - 384\tau_2^3 - 96\tau_2^2\tau_3 +
24192\tau_1^4 + 5856\tau_1^3\tau_2\,, $

$ A_  {38} =  1043821209600 + 407106518400\tau_1 +
125620772640\tau_2 + 34887476880\tau_3
   + 6552155040\tau_4 - 333453600\tau_5 - 45246680\tau_6 -
1717310\tau_7 + 20410\tau_8
   + 4402892160\tau_1^2 + 1411617888\tau_1\tau_2 +
963533088\tau_1\tau_3 + 182005662\tau_1\tau_4
   - 8641616\tau_1\tau_5 - 1279190\tau_1\tau_6 - 55998\tau_1\tau_7
+
460\tau_1\tau_8 - 143002176\tau_2^2
   - 31326592\tau_2\tau_3 - 6473770\tau_2\tau_4 +
110848\tau_2\tau_5
+ 6968\tau_2\tau_6 + 732\tau_2\tau_7
   - 48\tau_2\tau_8 + 1223280\tau_3^2 - 353012\tau_3\tau_4 -
39528\tau_3\tau_5 - 3198\tau_3\tau_6 - 182\tau_3\tau_7
   - 12\tau_3\tau_8 - 120192\tau_4^2 + 3156\tau_4\tau_5 +
744\tau_4\tau_6 + 9\tau_4\tau_7 + 136\tau_5^2 + 20\tau_5\tau_6
   + 7\tau_6^2 - 161638848\tau_1^3 - 49514112\tau_1^2\tau_2 -
3176352\tau_1^2\tau_3 - 37682\tau_1^2\tau_4
   + 42096\tau_1^2\tau_5 + 1996\tau_1^2\tau_6 - 224\tau_1^2\tau_7
-
716256\tau_1\tau_2^2 - 819424\tau_1\tau_2\tau_3
   - 2538\tau_1\tau_2\tau_4 - 816\tau_1\tau_2\tau_5 -
438\tau_1\tau_2\tau_6 - 8\tau_1\tau_2\tau_7 - 79560\tau_1\tau_3^2
   - 18940\tau_1\tau_3\tau_4 + 408\tau_1\tau_3\tau_5 +
80\tau_1\tau_3\tau_6 - 1104\tau_1\tau_4^2 + 20\tau_1\tau_4\tau_5 +
5\tau_1\tau_4\tau_6
   + 125184\tau_2^3 + 34400\tau_2^2\tau_3 + 4032\tau_2^2\tau_4 -
112\tau_2^2\tau_5 + 1472\tau_2\tau_3^2 + 470\tau_2\tau_3\tau_4
   + 7\tau_2\tau_4^2 + 1076544\tau_1^4 + 226752\tau_1^3\tau_2 +
22464\tau_1^3\tau_3 + 4032\tau_1^3\tau_4
   - 13152\tau_1^2\tau_2^2 + 1440\tau_1^2\tau_2\tau_3 +
320\tau_1^2\tau_2\tau_4 - 768\tau_1\tau_2^3 -
80\tau_1\tau_2^2\tau_3\,, $

$ A_  {44} =  - 111974400 - 40134528\tau_1 - 13098912\tau_2 -
3525264\tau_3 - 645408\tau_4
   + 31872\tau_5 + 3140\tau_6 + 206\tau_7 + 2\tau_8 +
483840\tau_1^2
+ 112704\tau_1\tau_2 - 6192\tau_1\tau_3
   - 158\tau_1\tau_4 - 208\tau_1\tau_5 - 56\tau_1\tau_6 -
4800\tau_2^2 - 2240\tau_2\tau_3 - 502\tau_2\tau_4 + 16\tau_2\tau_5
   - 144\tau_3^2 - 36\tau_3\tau_4 - 8\tau_4^2 - 192\tau_1^2\tau_2
+ 96\tau_1\tau_2^2\,, $

$ A_  {45} =  - 1077442560 - 339790464\tau_1 - 131923296\tau_2 -
36466416\tau_3 - 6904416\tau_4
   + 348192\tau_5 + 47160\tau_6 + 2058\tau_7 - 6\tau_8 +
25062912\tau_1^2 + 7864896\tau_1\tau_2
   + 1641456\tau_1\tau_3 + 295102\tau_1\tau_4 - 16208\tau_1\tau_5
-
2124\tau_1\tau_6 - 56\tau_1\tau_7 - 4608\tau_2^2
   - 14400\tau_2\tau_3 - 4254\tau_2\tau_4 + 224\tau_2\tau_5 -
2232\tau_3^2 - 428\tau_3\tau_4 + 32\tau_3\tau_5 + 5\tau_3\tau_6
   + 28\tau_4^2 - 9\tau_4\tau_5 - 241920\tau_1^3 -
56640\tau_1^2\tau_2 + 5184\tau_1^2\tau_3 + 980\tau_1^2\tau_4 +
96\tau_1\tau_2^2
   - 176\tau_1\tau_2\tau_3\,,
$

$ A_  {46} =  16940482560 + 5949821952\tau_1 + 2048224512\tau_2 +
567497088\tau_3
   + 107247360\tau_4 - 5441280\tau_5 - 746560\tau_6 - 27952\tau_7
+
368\tau_8
   - 168860160\tau_1^2 - 55602816\tau_1\tau_2 -
6568128\tau_1\tau_3
- 1031472\tau_1\tau_4
   + 68800\tau_1\tau_5 + 6568\tau_1\tau_6 + 152\tau_1\tau_7 -
1764672\tau_2^2 - 441440\tau_2\tau_3
   - 32176\tau_2\tau_4 + 544\tau_2\tau_5 - 212\tau_2\tau_6 +
4\tau_2\tau_7 - 7200\tau_3^2 - 496\tau_3\tau_4 - 224\tau_3\tau_5
   - 44\tau_3\tau_6 - 104\tau_4^2 - 10\tau_4\tau_6 +
1257984\tau_1^3
+ 324864\tau_1^2\tau_2 - 13824\tau_1^2\tau_3
   - 2408\tau_1^2\tau_4 + 6144\tau_1\tau_2^2 -
448\tau_1\tau_2\tau_3 - 92\tau_1\tau_2\tau_4 + 192\tau_2^3 +
32\tau_2^2\tau_3\,, $

$ A_  {47} =  - 87588864000 - 34068003840\tau_1 -
10510752000\tau_2 - 2881918080\tau_3
   - 546136320\tau_4 + 27768960\tau_5 + 3799120\tau_6 +
145360\tau_7
- 1280\tau_8
   - 338090112\tau_1^2 - 103373664\tau_1\tau_2 -
63507312\tau_1\tau_3 - 13851920\tau_1\tau_4
   + 640320\tau_1\tau_5 + 107284\tau_1\tau_6 + 4930\tau_1\tau_7 -
10\tau_1\tau_8 + 13986240\tau_2^2
   + 7844960\tau_2\tau_3 + 882320\tau_2\tau_4 - 29792\tau_2\tau_5
+
1356\tau_2\tau_6 - 44\tau_2\tau_7
   + 1164096\tau_3^2 + 325176\tau_3\tau_4 - 12640\tau_3\tau_5 -
632\tau_3\tau_6 - 44\tau_3\tau_7 + 20984\tau_4^2
   - 1576\tau_4\tau_5 - 73\tau_4\tau_6 - 12\tau_4\tau_7 +
32\tau_5^2
+ 4\tau_5\tau_6 + 14515200\tau_1^3
   + 4119744\tau_1^2\tau_2 + 186480\tau_1^2\tau_3 +
23486\tau_1^2\tau_4 - 5104\tau_1^2\tau_5 - 824\tau_1^2\tau_6
   - 30528\tau_1\tau_2^2 - 13440\tau_1\tau_2\tau_3 -
5534\tau_1\tau_2\tau_4 + 240\tau_1\tau_2\tau_5 +
4752\tau_1\tau_3^2
   + 968\tau_1\tau_3\tau_4 + 21\tau_1\tau_4^2 - 5184\tau_2^3 -
1952\tau_2^2\tau_3 - 224\tau_2\tau_3^2 - 96768\tau_1^4
   - 19776\tau_1^3\tau_2 + 1056\tau_1^2\tau_2^2\,,
$
\newpage
$ A_  {48} =  - 1687479091200 - 682571888640\tau_1 -
187352202240\tau_2 - 55304881920\tau_3
   - 10593477120\tau_4 + 544823040\tau_5 + 74649120\tau_6 +
2778240\tau_7 - 35520\tau_8
   - 16400012544\tau_1^2 + 537008832\tau_1\tau_2 -
1943112672\tau_1\tau_3 - 444557920\tau_1\tau_4
   + 23683904\tau_1\tau_5 + 3627632\tau_1\tau_6 +
131812\tau_1\tau_7
- 1804\tau_1\tau_8 + 2149212768\tau_2^2
   + 718934832\tau_2\tau_3 + 111819168\tau_2\tau_4 -
4638496\tau_2\tau_5 - 545744\tau_2\tau_6
   - 28058\tau_2\tau_7 + 182\tau_2\tau_8 + 36868320\tau_3^2 +
8592864\tau_3\tau_4 - 162208\tau_3\tau_5
   - 10016\tau_3\tau_6 - 1976\tau_3\tau_7 - 12\tau_3\tau_8 +
339048\tau_4^2 + 13072\tau_4\tau_5 + 3769\tau_4\tau_6
   - 142\tau_4\tau_7 - 15\tau_4\tau_8 - 1504\tau_5^2 -
508\tau_5\tau_6 - 63\tau_6^2 + 3\tau_6\tau_7 + 226533888\tau_1^3
   - 14490240\tau_1^2\tau_2 - 11316960\tau_1^2\tau_3 -
3379668\tau_1^2\tau_4 + 73440\tau_1^2\tau_5
   + 19408\tau_1^2\tau_6 + 2080\tau_1^2\tau_7 -
20720064\tau_1\tau_2^2 + 2421552\tau_1\tau_2\tau_3
   + 152166\tau_1\tau_2\tau_4 + 14704\tau_1\tau_2\tau_5 +
2232\tau_1\tau_2\tau_6 + 96\tau_1\tau_2\tau_7 +
700416\tau_1\tau_3^2
   + 156600\tau_1\tau_3\tau_4 - 3232\tau_1\tau_3\tau_5 -
616\tau_1\tau_3\tau_6 + 5800\tau_1\tau_4^2 - 88\tau_1\tau_4\tau_5
   - 45\tau_1\tau_4\tau_6 - 807936\tau_2^3 - 185248\tau_2^2\tau_3
-
3058\tau_2^2\tau_4 + 48\tau_2^2\tau_5 + 2800\tau_2\tau_3^2
   + 1656\tau_2\tau_3\tau_4 - 96\tau_2\tau_3\tau_5 +
42\tau_2\tau_4^2 + 864\tau_3^3 + 216\tau_3^2\tau_4 +
15\tau_3\tau_4^2
   - 870912\tau_1^4 + 234624\tau_1^3\tau_2 - 207360\tau_1^3\tau_3
-
35392\tau_1^3\tau_4 + 123264\tau_1^2\tau_2^2
   - 15104\tau_1^2\tau_2\tau_3 - 2720\tau_1^2\tau_2\tau_4 +
2592\tau_1\tau_2^3 + 320\tau_1\tau_2^2\tau_3\,, $

$ A_  {55} = 20048394240 + 8292917376\tau_1 + 2441404512\tau_2 +
660526704\tau_3
   + 125330784\tau_4 - 6307968\tau_5 - 860244\tau_6 - 35010\tau_7
+
138\tau_8
   + 259687296\tau_1^2 + 96864000\tau_1\tau_2 +
31537008\tau_1\tau_3
+ 6466434\tau_1\tau_4
   - 296112\tau_1\tau_5 - 43952\tau_1\tau_6 - 1868\tau_1\tau_7 +
4\tau_1\tau_8 + 1043712\tau_2^2
   - 512928\tau_2\tau_3 + 70114\tau_2\tau_4 - 1824\tau_2\tau_5 -
920\tau_2\tau_6 - 40\tau_2\tau_7 - 230112\tau_3^2
   - 49732\tau_3\tau_4 + 2864\tau_3\tau_5 + 312\tau_3\tau_6 +
2\tau_3\tau_7 - 1032\tau_4^2 + 150\tau_4\tau_5 + 6\tau_4\tau_6
   - 12\tau_5^2 - 5346432\tau_1^3 - 1691520\tau_1^2\tau_2 +
18720\tau_1^2\tau_3 + 9524\tau_1^2\tau_4 + 784\tau_1^2\tau_5
   + 32\tau_1^2\tau_6 - 66432\tau_1\tau_2^2 -
3328\tau_1\tau_2\tau_3
+ 724\tau_1\tau_2\tau_4 - 8\tau_1\tau_2\tau_5 - 864\tau_1\tau_3^2
   - 122\tau_1\tau_3\tau_4 - 384\tau_2^3 - 96\tau_2^2\tau_3 +
20\tau_2\tau_3^2 + 24192\tau_1^4 + 4896\tau_1^3\tau_2\,, $

$ A_ {56} =  45013708800 + 12445726464\tau_1 + 5402781504\tau_2 +
1524340512\tau_3
   + 284556096\tau_4 - 14496000\tau_5 - 1974432\tau_6 -
71580\tau_7
+ 1188\tau_8
   - 1650820608\tau_1^2 - 564132864\tau_1\tau_2 -
122574240\tau_1\tau_3 - 23833652\tau_1\tau_4
   + 1237504\tau_1\tau_5 + 162020\tau_1\tau_6 + 6480\tau_1\tau_7 -
52\tau_1\tau_8 - 8246208\tau_2^2
   + 212688\tau_2\tau_3 - 215108\tau_2\tau_4 + 4864\tau_2\tau_5 -
352\tau_2\tau_6 + 180\tau_2\tau_7 + 622296\tau_3^2
   + 130724\tau_3\tau_4 - 7560\tau_3\tau_5 - 919\tau_3\tau_6 -
5\tau_3\tau_7 + 1784\tau_4^2 - 160\tau_4\tau_5
   + 11\tau_4\tau_6 - 12\tau_5\tau_6 + 27385344\tau_1^3 +
8846976\tau_1^2\tau_2 + 359712\tau_1^2\tau_3
   + 37200\tau_1^2\tau_4 - 5920\tau_1^2\tau_5 - 320\tau_1^2\tau_6
+
391296\tau_1\tau_2^2 + 62528\tau_1\tau_2\tau_3
   + 3348\tau_1\tau_2\tau_4 - 192\tau_1\tau_2\tau_5 +
5184\tau_1\tau_3^2 + 773\tau_1\tau_3\tau_4 + 7\tau_1\tau_4^2 +
5568\tau_2^3
   + 992\tau_2^2\tau_3 - 32\tau_2\tau_3^2 + 5\tau_2\tau_3\tau_4 -
145152\tau_1^4 - 36864\tau_1^3\tau_2 - 960\tau_1^2\tau_2^2\,, 
$
\newpage
$
 A_  {57} =  1246623436800 + 483052066560\tau_1 + 148432852800\tau_2
+
42092848800\tau_3
   + 7820301120\tau_4 - 397619520\tau_5 - 53694480\tau_6 -
2049900\tau_7 + 22980\tau_8
   + 4300829568\tau_1^2 + 760490784\tau_1\tau_2 +
1215098928\tau_1\tau_3 + 200005628\tau_1\tau_4
   - 9333280\tau_1\tau_5 - 1332604\tau_1\tau_6 - 60062\tau_1\tau_7
+
614\tau_1\tau_8 - 355268736\tau_2^2
   - 31709952\tau_2\tau_3 - 16452340\tau_2\tau_4 +
595408\tau_2\tau_5 + 85556\tau_2\tau_6 + 4624\tau_2\tau_7
   - 48\tau_2\tau_8 + 17004960\tau_3^2 + 2761260\tau_3\tau_4 -
205216\tau_3\tau_5 - 23297\tau_3\tau_6
   - 452\tau_3\tau_7 + 3\tau_3\tau_8 - 91656\tau_4^2 -
1588\tau_4\tau_5 + 251\tau_4\tau_6 + 109\tau_4\tau_7 + 296\tau_5^2
   + 87\tau_5\tau_6 - 15\tau_5\tau_7 + 7\tau_6^2 -
121830912\tau_1^3
- 29644800\tau_1^2\tau_2 - 1571472\tau_1^2\tau_3
   + 1010070\tau_1^2\tau_4 - 4560\tau_1^2\tau_5 -
6776\tau_1^2\tau_6
- 596\tau_1^2\tau_7 + 696960\tau_1\tau_2^2
   - 2064576\tau_1\tau_2\tau_3 - 6966\tau_1\tau_2\tau_4 -
2080\tau_1\tau_2\tau_5 - 1008\tau_1\tau_2\tau_6 -
32\tau_1\tau_2\tau_7
   - 155376\tau_1\tau_3^2 - 28776\tau_1\tau_3\tau_4 +
1400\tau_1\tau_3\tau_5 + 131\tau_1\tau_3\tau_6 -
1491\tau_1\tau_4^2
   + 30\tau_1\tau_4\tau_5 + 5\tau_1\tau_4\tau_6 + 98688\tau_2^3 -
7968\tau_2^2\tau_3 + 4032\tau_2^2\tau_4 - 112\tau_2^2\tau_5
   - 9296\tau_2\tau_3^2 + 164\tau_2\tau_3\tau_4 +
16\tau_2\tau_3\tau_5 + 7\tau_2\tau_4^2 - 432\tau_3^3 -
51\tau_3^2\tau_4
   + 677376\tau_1^4 + 91968\tau_1^3\tau_2 + 62208\tau_1^3\tau_3 +
8288\tau_1^3\tau_4 - 19104\tau_1^2\tau_2^2
   + 6256\tau_1^2\tau_2\tau_3 + 560\tau_1^2\tau_2\tau_4 -
768\tau_1\tau_2^3 - 80\tau_1\tau_2^2\tau_3\,, $

$ A_  {58} =  10620274176000 + 4522739639040\tau_1 +
1343299199040\tau_2
   + 376374552480\tau_3 + 65097330240\tau_4 - 3232884480\tau_5 -
426517680\tau_6
   - 17487180\tau_7 + 184740\tau_8 + 194726218752\tau_1^2 +
85748180736\tau_1\tau_2
   + 30808137312\tau_1\tau_3 + 3685884732\tau_1\tau_4 -
151976256\tau_1\tau_5 - 17286964\tau_1\tau_6
   - 1203544\tau_1\tau_7 + 8924\tau_1\tau_8 + 6609626592\tau_2^2 +
4563278928\tau_2\tau_3
   + 179570316\tau_2\tau_4 - 1500656\tau_2\tau_5 +
1027948\tau_2\tau_6 - 125738\tau_2\tau_7 + 782\tau_2\tau_8
   + 750489264\tau_3^2 + 91775452\tau_3\tau_4 -
2606528\tau_3\tau_5
- 13827\tau_3\tau_6 - 48782\tau_3\tau_7
   + 735\tau_3\tau_8 - 9323128\tau_4^2 + 1337684\tau_4\tau_5 +
239281\tau_4\tau_6 - 483\tau_4\tau_7 + 81\tau_4\tau_8
   - 43800\tau_5^2 - 14877\tau_5\tau_6 - 85\tau_5\tau_7 -
18\tau_5\tau_8 - 1136\tau_6^2 + 19\tau_6\tau_7
   + 244871424\tau_1^3 + 345814080\tau_1^2\tau_2 +
366058944\tau_1^2\tau_3 + 71736920\tau_1^2\tau_4
   - 3315648\tau_1^2\tau_5 - 477028\tau_1^2\tau_6 -
11452\tau_1^2\tau_7 + 184\tau_1^2\tau_8 + 39638592\tau_1\tau_2^2
   + 44512432\tau_1\tau_2\tau_3 + 7169834\tau_1\tau_2\tau_4 -
387280\tau_1\tau_2\tau_5 - 39204\tau_1\tau_2\tau_6
   + 1520\tau_1\tau_2\tau_7 - 40\tau_1\tau_2\tau_8 +
12756168\tau_1\tau_3^2 + 2269382\tau_1\tau_3\tau_4
   - 101912\tau_1\tau_3\tau_5 - 10942\tau_1\tau_3\tau_6 -
484\tau_1\tau_3\tau_7 - 52914\tau_1\tau_4^2 +
997\tau_1\tau_4\tau_5
   + 566\tau_1\tau_4\tau_6 + 4\tau_1\tau_4\tau_7 +
176\tau_1\tau_5^2
+ 24\tau_1\tau_5\tau_6 + 6\tau_1\tau_6^2 + 5101056\tau_2^3
   + 1963424\tau_2^2\tau_3 + 345294\tau_2^2\tau_4 -
2128\tau_2^2\tau_5 + 1952\tau_2^2\tau_6 + 96\tau_2^2\tau_7
   + 129928\tau_2\tau_3^2 + 122262\tau_2\tau_3\tau_4 -
2448\tau_2\tau_3\tau_5 - 205\tau_2\tau_3\tau_6 -
4\tau_2\tau_3\tau_7
   + 7131\tau_2\tau_4^2 - 239\tau_2\tau_4\tau_5 -
24\tau_2\tau_4\tau_6 - 34560\tau_3^3 - 6609\tau_3^2\tau_4 +
384\tau_3^2\tau_5
   + 78\tau_3^2\tau_6 + 59\tau_3\tau_4^2 + 10\tau_3\tau_4\tau_5 +
4\tau_3\tau_4\tau_6 + 8\tau_4^3 - 98751744\tau_1^4
   - 36422784\tau_1^3\tau_2 - 6322176\tau_1^3\tau_3 -
776660\tau_1^3\tau_4 + 48064\tau_1^3\tau_5 + 4256\tau_1^3\tau_6
   - 1445184\tau_1^2\tau_2^2 - 667776\tau_1^2\tau_2\tau_3 -
25900\tau_1^2\tau_2\tau_4 + 128\tau_1^2\tau_2\tau_5 -
224\tau_1^2\tau_2\tau_6
   + 10368\tau_1^2\tau_3^2 + 508\tau_1^2\tau_3\tau_4 -
476\tau_1^2\tau_4^2 + 100512\tau_1\tau_2^3 +
12640\tau_1\tau_2^2\tau_3
   + 1080\tau_1\tau_2^2\tau_4 - 96\tau_1\tau_2^2\tau_5 +
1856\tau_1\tau_2\tau_3^2 + 624\tau_1\tau_2\tau_3\tau_4 +
6\tau_1\tau_2\tau_4^2
   + 1536\tau_2^4 + 128\tau_2^3\tau_3 - 64\tau_2^2\tau_3^2 +
870912\tau_1^5 + 200448\tau_1^4\tau_2 - 5760\tau_1^3\tau_2^2 -
   384\tau_1^2\tau_2^3\,,
$
\newpage
$ A_  {66} =  258626027520 + 108776079360\tau_1 +
31289152512\tau_2 + 8572294656\tau_3
   + 1620636672\tau_4 - 82225152\tau_5 - 11177152\tau_6 -
429376\tau_7 + 5120\tau_8
   + 3883417344\tau_1^2 + 1356812736\tau_1\tau_2 +
476416800\tau_1\tau_3 + 93942272\tau_1\tau_4
   - 4529280\tau_1\tau_5 - 641800\tau_1\tau_6 - 28252\tau_1\tau_7
+
316\tau_1\tau_8 - 17069472\tau_2^2
   - 11507984\tau_2\tau_3 - 1022752\tau_2\tau_4 +
24192\tau_2\tau_5
+ 3332\tau_2\tau_6 - 466\tau_2\tau_7
   + 2\tau_2\tau_8 - 2128320\tau_3^2 - 647328\tau_3\tau_4 +
24832\tau_3\tau_5 + 4184\tau_3\tau_6 + 8\tau_3\tau_7
   - 48800\tau_4^2 + 3744\tau_4\tau_5 + 652\tau_4\tau_6 +
4\tau_4\tau_7 - 64\tau_5^2 - 24\tau_5\tau_6 - 14\tau_6^2
   - 111186432\tau_1^3 - 37928064\tau_1^2\tau_2 -
2740896\tau_1^2\tau_3 - 310340\tau_1^2\tau_4
   + 33952\tau_1^2\tau_5 + 2192\tau_1^2\tau_6 -
1877568\tau_1\tau_2^2 - 426672\tau_1\tau_2\tau_3 -
21682\tau_1\tau_2\tau_4
   + 784\tau_1\tau_2\tau_5 - 152\tau_1\tau_2\tau_6 -
25056\tau_1\tau_3^2 - 4304\tau_1\tau_3\tau_4 - 126\tau_1\tau_4^2
   - 27840\tau_2^3 - 7232\tau_2^2\tau_3 - 502\tau_2^2\tau_4 +
16\tau_2^2\tau_5 - 208\tau_2\tau_3^2 - 36\tau_2\tau_3\tau_4
   + 774144\tau_1^4 + 206208\tau_1^3\tau_2 + 5760\tau_1^2\tau_2^2
+ 96\tau_1\tau_2^3\,, $

$ A_  {67} =  - 54942105600 - 16578017280\tau_1 - 2812730880\tau_2
- 3529140480\tau_3
   - 158860800\tau_4 + 18343680\tau_5 + 4894720\tau_6 +
83680\tau_7
+ 1120\tau_8
   + 827442432\tau_1^2 + 1886712384\tau_1\tau_2 -
476402976\tau_1\tau_3 + 88752992\tau_1\tau_4
   - 808064\tau_1\tau_5 + 778040\tau_1\tau_6 - 5076\tau_1\tau_7 +
324\tau_1\tau_8 + 468574176\tau_2^2
   - 75387184\tau_2\tau_3 + 46788208\tau_2\tau_4 -
1124224\tau_2\tau_5 + 143076\tau_2\tau_6
   - 7194\tau_2\tau_7 + 42\tau_2\tau_8 - 56272176\tau_3^2 -
4241016\tau_3\tau_4 + 555840\tau_3\tau_5
   + 154668\tau_3\tau_6 + 2586\tau_3\tau_7 - 42\tau_3\tau_8 +
1216144\tau_4^2 - 59488\tau_4\tau_5 + 6321\tau_4\tau_6
   - 329\tau_4\tau_7 - 64\tau_5^2 - 728\tau_5\tau_6 - 98\tau_6^2 -
16\tau_6\tau_7 - 275304960\tau_1^3
   - 111598848\tau_1^2\tau_2 - 3884832\tau_1^2\tau_3 -
5361964\tau_1^2\tau_4 + 228896\tau_1^2\tau_5
   + 20712\tau_1^2\tau_6 + 1736\tau_1^2\tau_7 -
5754624\tau_1\tau_2^2 + 6490800\tau_1\tau_2\tau_3
   - 332978\tau_1\tau_2\tau_4 + 10096\tau_1\tau_2\tau_5 -
788\tau_1\tau_2\tau_6 + 140\tau_1\tau_2\tau_7 +
950256\tau_1\tau_3^2
   + 171398\tau_1\tau_3\tau_4 - 8912\tau_1\tau_3\tau_5 -
716\tau_1\tau_3\tau_6 + 6559\tau_1\tau_4^2 - 72\tau_1\tau_4\tau_5
   + 4\tau_1\tau_4\tau_6 + 121152\tau_2^3 + 298752\tau_2^2\tau_3 -
4542\tau_2^2\tau_4 - 112\tau_2^2\tau_5 + 79728\tau_2\tau_3^2
   + 4002\tau_2\tau_3\tau_4 - 176\tau_2\tau_3\tau_5 -
7\tau_2\tau_4^2 + 4\tau_2\tau_4\tau_5 + 3888\tau_3^3 +
528\tau_3^2\tau_4
   + 6\tau_3\tau_4^2 + 2612736\tau_1^4 + 796032\tau_1^3\tau_2
-283392\tau_1^3\tau_3 - 23576\tau_1^3\tau_4
   + 49920\tau_1^2\tau_2^2 - 48640\tau_1^2\tau_2\tau_3 -
2436\tau_1^2\tau_2\tau_4 + 2592\tau_1\tau_2^3\,, $

$ A_  {68} =  - 12499528089600 - 5461733836800\tau_1 -
1318777459200\tau_2
   - 333591609600\tau_3 - 41745477120\tau_4 + 1309455360\tau_5 +
249137280\tau_6
   + 16727520\tau_7 - 331680\tau_8 - 256398974208\tau_1^2 -
22213180992\tau_1\tau_2
   - 241790688\tau_1\tau_3 + 7563312032\tau_1\tau_4 -
677547904\tau_1\tau_5 - 65199624\tau_1\tau_6
   + 260372\tau_1\tau_7 - 47508\tau_1\tau_8 + 24140669472\tau_2^2
+
16753774320\tau_2\tau_3
   + 5681682544\tau_2\tau_4 - 384786688\tau_2\tau_5 -
43575284\tau_2\tau_6 - 773110\tau_2\tau_7
   - 5594\tau_2\tau_8 + 2801047248\tau_3^2 +
1758229880\tau_3\tau_4
- 115857536\tau_3\tau_5
   - 13380236\tau_3\tau_6 - 260958\tau_3\tau_7 - 858\tau_3\tau_8 +
232300496\tau_4^2 - 28617088\tau_4\tau_5
   - 3437322\tau_4\tau_6 - 83953\tau_4\tau_7 + 269\tau_4\tau_8 +
853632\tau_5^2 + 208880\tau_5\tau_6
   + 5640\tau_5\tau_7 - 32\tau_5\tau_8 + 12339\tau_6^2 +
678\tau_6\tau_7 - 20\tau_6\tau_8 + 5\tau_7^2
   + 8438404608\tau_1^3 + 998146176\tau_1^2\tau_2 -
23184288\tau_1^2\tau_3 - 114754100\tau_1^2\tau_4
   + 13495200\tau_1^2\tau_5 + 1406864\tau_1^2\tau_6 +
20528\tau_1^2\tau_7 + 560\tau_1^2\tau_8 - 496877376\tau_1\tau_2^2
   - 151100944\tau_1\tau_2\tau_3 - 61896062\tau_1\tau_2\tau_4 +
6241008\tau_1\tau_2\tau_5 + 647224\tau_1\tau_2\tau_6
   + 10616\tau_1\tau_2\tau_7 + 56\tau_1\tau_2\tau_8 +
12808656\tau_1\tau_3^2 + 5836846\tau_1\tau_3\tau_4
   + 206608\tau_1\tau_3\tau_5 + 4296\tau_1\tau_3\tau_6 +
184\tau_1\tau_3\tau_7 + 937287\tau_1\tau_4^2 +
11144\tau_1\tau_4\tau_5
   - 3012\tau_1\tau_4\tau_6 - 240\tau_1\tau_4\tau_7 -
3840\tau_1\tau_5^2 - 704\tau_1\tau_5\tau_6 - 64\tau_1\tau_6^2
   - 16784640\tau_2^3 - 10005024\tau_2^2\tau_3 -
3321730\tau_2^2\tau_4 + 235216\tau_2^2\tau_5
   + 16388\tau_2^2\tau_6 + 204\tau_2^2\tau_7 -
712464\tau_2\tau_3^2
- 878654\tau_2\tau_3\tau_4 + 64368\tau_2\tau_3\tau_5
   + 3332\tau_2\tau_3\tau_6 + 104\tau_2\tau_3\tau_7 -
54891\tau_2\tau_4^2 + 5940\tau_2\tau_4\tau_5 +
70\tau_2\tau_4\tau_6
   + 3\tau_2\tau_4\tau_7 - 128\tau_2\tau_5^2 + 195696\tau_3^3 +
18120\tau_3^2\tau_4 + 192\tau_3^2\tau_5 - 312\tau_3^2\tau_6
   - 3059\tau_3\tau_4^2 + 320\tau_3\tau_4\tau_5 -
\tau_3\tau_4\tau_6
- 68\tau_4^3 + 5\tau_4^2\tau_5 - 98025984\tau_1^4
   - 21462144\tau_1^3\tau_2 - 2618496\tau_1^3\tau_3 -
163376\tau_1^3\tau_4 - 62848\tau_1^3\tau_5 - 5696\tau_1^3\tau_6
   + 1489152\tau_1^2\tau_2^2 - 908288\tau_1^2\tau_2\tau_3 -
103176\tau_1^2\tau_2\tau_4 - 17472\tau_1^2\tau_2\tau_5
   - 1184\tau_1^2\tau_2\tau_6 - 65664\tau_1^2\tau_3^2 +
6344\tau_1^2\tau_3\tau_4 + 3087\tau_1^2\tau_4^2 +
220320\tau_1\tau_2^3
   + 39712\tau_1\tau_2^2\tau_3 + 1188\tau_1\tau_2^2\tau_4 -
576\tau_1\tau_2^2\tau_5 - 14144\tau_1\tau_2\tau_3^2 -
2976\tau_1\tau_2\tau_3\tau_4
   - 9\tau_1\tau_2\tau_4^2 + 6720\tau_2^4 + 4064\tau_2^3\tau_3 +
576\tau_2^2\tau_3^2 + 387072\tau_1^5 + 152064\tau_1^4\tau_2
   + 15360\tau_1^3\tau_2^2 - 384\tau_1^2\tau_2^3\,,
$

$ A_  {77} =  16780433817600 + 7902132940800\tau_1 +
2437786859520\tau_2
   + 635194874880\tau_3 + 103944069120\tau_4 - 4777889280\tau_5 -
671874880\tau_6
   - 23098240\tau_7 + 322880\tau_8 + 546982951680\tau_1^2 +
337080365760\tau_1\tau_2
   + 88153725600\tau_1\tau_3 + 11017394880\tau_1\tau_4 -
365732480\tau_1\tau_5 - 59534120\tau_1\tau_6
   - 1507276\tau_1\tau_7 + 37548\tau_1\tau_8 + 48334917600\tau_2^2
+
22589290480\tau_2\tau_3
   + 2403174400\tau_2\tau_4 - 63536160\tau_2\tau_5 -
11877260\tau_2\tau_6 - 158682\tau_2\tau_7
   + 8778\tau_2\tau_8 + 2535608160\tau_3^2 + 435835280\tau_3\tau_4
-
6047120\tau_3\tau_5
   - 1711520\tau_3\tau_6 + 12828\tau_3\tau_7 + 1884\tau_3\tau_8 -
7950720\tau_4^2 + 3850320\tau_4\tau_5
   + 344270\tau_4\tau_6 + 27502\tau_4\tau_7 + 76\tau_4\tau_8 -
174400\tau_5^2 - 37860\tau_5\tau_6 - 2210\tau_5\tau_7
   + 2\tau_5\tau_8 - 1850\tau_6^2 - 216\tau_6\tau_7 - 20\tau_7^2 -
16339059072\tau_1^3 - 8319935136\tau_1^2\tau_2
   - 984666096\tau_1^2\tau_3 - 92044084\tau_1^2\tau_4 +
3055648\tau_1^2\tau_5 + 652724\tau_1^2\tau_6
   + 870\tau_1^2\tau_7 - 166\tau_1^2\tau_8 -
1147642944\tau_1\tau_2^2 - 230924656\tau_1\tau_2\tau_3
   - 10252570\tau_1\tau_2\tau_4 - 250800\tau_1\tau_2\tau_5 +
54176\tau_1\tau_2\tau_6 - 3328\tau_1\tau_2\tau_7
   - 40\tau_1\tau_2\tau_8 + 2959488\tau_1\tau_3^2 +
1688412\tau_1\tau_3\tau_4 - 114896\tau_1\tau_3\tau_5
   - 1744\tau_1\tau_3\tau_6 - 376\tau_1\tau_3\tau_7 +
71464\tau_1\tau_4^2 - 22430\tau_1\tau_4\tau_5 -
88\tau_1\tau_4\tau_6
   + 78\tau_1\tau_4\tau_7 + 1520\tau_1\tau_5^2 +
224\tau_1\tau_5\tau_6 + 6\tau_1\tau_6^2 - 43939776\tau_2^3
   - 19313216\tau_2^2\tau_3 - 1153262\tau_2^2\tau_4 +
10768\tau_2^2\tau_5 + 1952\tau_2^2\tau_6 + 96\tau_2^2\tau_7
   - 2487504\tau_2\tau_3^2 - 207480\tau_2\tau_3\tau_4 -
6624\tau_2\tau_3\tau_5 - 680\tau_2\tau_3\tau_6 -
24\tau_2\tau_3\tau_7
   + 2658\tau_2\tau_4^2 - 766\tau_2\tau_4\tau_5 -
24\tau_2\tau_4\tau_6 + 16\tau_2\tau_5^2 - 119520\tau_3^3 -
25952\tau_3^2\tau_4
   - 208\tau_3^2\tau_5 + 8\tau_3^2\tau_6 - 1140\tau_3\tau_4^2 -
36\tau_3\tau_4\tau_5 + 4\tau_3\tau_4\tau_6 + 8\tau_4^3
   + 130733568\tau_1^4 + 58016448\tau_1^3\tau_2 +
1294992\tau_1^3\tau_3 - 25878\tau_1^3\tau_4 + 9584\tau_1^3\tau_5
   - 2744\tau_1^3\tau_6 + 7370688\tau_1^2\tau_2^2 +
657152\tau_1^2\tau_2\tau_3 + 19986\tau_1^2\tau_2\tau_4 +
4048\tau_1^2\tau_2\tau_5
   - 224\tau_1^2\tau_2\tau_6 + 66096\tau_1^2\tau_3^2 +
13492\tau_1^2\tau_3\tau_4 - 70\tau_1^2\tau_4^2 +
285024\tau_1\tau_2^3
   + 28288\tau_1\tau_2^2\tau_3 + 1080\tau_1\tau_2^2\tau_4 -
96\tau_1\tau_2^2\tau_5 + 512\tau_1\tau_2\tau_3^2 +
824\tau_1\tau_2\tau_3\tau_4
   + 6\tau_1\tau_2\tau_4^2 + 1536\tau_2^4 + 128\tau_2^3\tau_3 -
64\tau_2^2\tau_3^2 - 193536\tau_1^5 - 108480\tau_1^4\tau_2
   - 13344\tau_1^3\tau_2^2 - 384\tau_1^2\tau_2^3\,,
$

$ A_  {78} =  285552635904000 + 174924574617600\tau_1 +
53238191063040\tau_2
   + 12505022645760\tau_3 + 2073096714240\tau_4 -
83668565760\tau_5
   - 17581805760\tau_6 - 601907520\tau_7 + 9037440\tau_8 +
23905643823360\tau_1^2
   + 15121118690880\tau_1\tau_2 + 3484047126240\tau_1\tau_3 +
553054514560\tau_1\tau_4
   - 19830621440\tau_1\tau_5 - 4952932360\tau_1\tau_6 -
168074052\tau_1\tau_7 + 2601236\tau_1\tau_8
   + 2261698962720\tau_2^2 + 982106623440\tau_2\tau_3 +
151266515280\tau_2\tau_4
   - 5096249760\tau_2\tau_5 - 1446441140\tau_2\tau_6 -
46848614\tau_2\tau_7 + 757286\tau_2\tau_8
   + 99487163520\tau_3^2 + 27947465480\tau_3\tau_4 -
708851280\tau_3\tau_5 - 304226800\tau_3\tau_6
   - 9279224\tau_3\tau_7 + 162888\tau_3\tau_8 + 1736990800\tau_4^2
-
41798000\tau_4\tau_5
   - 45009765\tau_4\tau_6 - 1279301\tau_4\tau_7 +
24302\tau_4\tau_8
- 2353600\tau_5^2 + 1352180\tau_5\tau_6
   + 29550\tau_5\tau_7 - 806\tau_5\tau_8 + 227595\tau_6^2 +
14698\tau_6\tau_7 - 200\tau_6\tau_8 + 185\tau_7^2
   - 24\tau_7\tau_8 - 532970654976\tau_1^3 -
200592241344\tau_1^2\tau_2 - 20629111104\tau_1^2\tau_3
   - 2797414908\tau_1^2\tau_4 + 149510816\tau_1^2\tau_5 +
58714928\tau_1^2\tau_6 + 634804\tau_1^2\tau_7
   - 23948\tau_1^2\tau_8 - 6480704928\tau_1\tau_2^2 +
4021005824\tau_1\tau_2\tau_3 + 1137929690\tau_1\tau_2\tau_4
   - 62664048\tau_1\tau_2\tau_5 + 4342872\tau_1\tau_2\tau_6 -
335182\tau_1\tau_2\tau_7 - 862\tau_1\tau_2\tau_8
   + 842121792\tau_1\tau_3^2 + 209925200\tau_1\tau_3\tau_4 -
4855248\tau_1\tau_3\tau_5 + 170640\tau_1\tau_3\tau_6
   - 69264\tau_1\tau_3\tau_7 + 588\tau_1\tau_3\tau_8 +
9028647\tau_1\tau_4^2 + 324226\tau_1\tau_4\tau_5
   + 145118\tau_1\tau_4\tau_6 - 5983\tau_1\tau_4\tau_7 +
57\tau_1\tau_4\tau_8 - 20080\tau_1\tau_5^2 -
20228\tau_1\tau_5\tau_6
   - 320\tau_1\tau_5\tau_7 - 1727\tau_1\tau_6^2 +
13\tau_1\tau_6\tau_7 + 925542912\tau_2^3 + 359682272\tau_2^2\tau_3
   + 107379998\tau_2^2\tau_4 - 8076720\tau_2^2\tau_5 -
504476\tau_2^2\tau_6 - 31892\tau_2^2\tau_7 + 112\tau_2^2\tau_8
   + 42374608\tau_2\tau_3^2 + 10971580\tau_2\tau_3\tau_4 -
1681760\tau_2\tau_3\tau_5 - 8920\tau_2\tau_3\tau_6
   - 3492\tau_2\tau_3\tau_7 - 32\tau_2\tau_3\tau_8 +
547457\tau_2\tau_4^2 - 123078\tau_2\tau_4\tau_5 +
4037\tau_2\tau_4\tau_6
   - 489\tau_2\tau_4\tau_7 + 2800\tau_2\tau_5^2 -
388\tau_2\tau_5\tau_6 - 28\tau_2\tau_6^2 + 6001920\tau_3^3
   + 701776\tau_3^2\tau_4 - 171280\tau_3^2\tau_5 -
9400\tau_3^2\tau_6 - 312\tau_3^2\tau_7 - 170838\tau_3\tau_4^2
   - 12316\tau_3\tau_4\tau_5 + 703\tau_3\tau_4\tau_6 -
\tau_3\tau_4\tau_7 + 576\tau_3\tau_5^2 + 104\tau_3\tau_5\tau_6 +
5\tau_3\tau_6^2
   - 13616\tau_4^3 + 518\tau_4^2\tau_5 + 87\tau_4^2\tau_6 +
3\tau_4\tau_5\tau_6 + 2356978176\tau_1^4
   - 539559936\tau_1^3\tau_2 - 171097056\tau_1^3\tau_3 -
21752484\tau_1^3\tau_4 + 1147616\tau_1^3\tau_5
   - 88048\tau_1^3\tau_6 + 11008\tau_1^3\tau_7 -
495148224\tau_1^2\tau_2^2 - 61453552\tau_1^2\tau_2\tau_3
   - 4322854\tau_1^2\tau_2\tau_4 + 589136\tau_1^2\tau_2\tau_5 -
11784\tau_1^2\tau_2\tau_6 + 992\tau_1^2\tau_2\tau_7
   - 127296\tau_1^2\tau_3^2 + 580792\tau_1^2\tau_3\tau_4 +
60704\tau_1^2\tau_3\tau_5 + 3752\tau_1^2\tau_3\tau_6 +
53548\tau_1^2\tau_4^2
   + 5272\tau_1^2\tau_4\tau_5 + 153\tau_1^2\tau_4\tau_6 -
54753120\tau_1\tau_2^3 - 9911264\tau_1\tau_2^2\tau_3
   - 594250\tau_1\tau_2^2\tau_4 + 47792\tau_1\tau_2^2\tau_5 +
1200\tau_1\tau_2^2\tau_6 + 80\tau_1\tau_2^2\tau_7 -
378288\tau_1\tau_2\tau_3^2
   + 103564\tau_1\tau_2\tau_3\tau_4 - 608\tau_1\tau_2\tau_3\tau_5
-
288\tau_1\tau_2\tau_3\tau_6 + 11426\tau_1\tau_2\tau_4^2
   - 64\tau_1\tau_2\tau_4\tau_5 - 20\tau_1\tau_2\tau_4\tau_6 +
7776\tau_1\tau_3^3 + 1800\tau_1\tau_3^2\tau_4 -
74\tau_1\tau_3\tau_4^2 + 7\tau_1\tau_4^3
   - 1043136\tau_2^4 - 265696\tau_2^3\tau_3 - 18896\tau_2^3\tau_4
+
256\tau_2^3\tau_5 - 4384\tau_2^2\tau_3^2
   + 288\tau_2^2\tau_3\tau_4 - 128\tau_2^2\tau_3\tau_5 -
28\tau_2^2\tau_4^2 + 192\tau_2\tau_3^3 + 320\tau_2\tau_3^2\tau_4 +
5\tau_2\tau_3\tau_4^2
   + 6289920\tau_1^5 + 11750016\tau_1^4\tau_2 -
276480\tau_1^4\tau_3
- 166432\tau_1^4\tau_4
   + 3310080\tau_1^3\tau_2^2 + 71680\tau_1^3\tau_2\tau_3 -
19360\tau_1^3\tau_2\tau_4 + 201312\tau_1^2\tau_2^3
   - 11072\tau_1^2\tau_2^2\tau_3 - 1968\tau_1^2\tau_2^2\tau_4 +
3072\tau_1\tau_2^4\,, $

$ A_  {88} =  7220703721881600 + 5289666799104000\tau_1 +
1785673491425280\tau_2
   + 496119491205120\tau_3 + 95576912148480\tau_4 -
5052882155520\tau_5
   - 885870784320\tau_6 - 5690661120\tau_7 - 359108160\tau_8 +
872136774224640\tau_1^2
   + 621176439456960\tau_1\tau_2 + 178976145612960\tau_1\tau_3 +
34704473400000\tau_1\tau_4
   - 1822764226560\tau_1\tau_5 - 325893113960\tau_1\tau_6 -
2228481164\tau_1\tau_7
   - 128098388\tau_1\tau_8 + 108917055491040\tau_2^2 +
60782999618160\tau_2\tau_3
   + 11754323132000\tau_2\tau_4 - 623141372960\tau_2\tau_5 -
109113405100\tau_2\tau_6
   - 768343898\tau_2\tau_7 - 42696758\tau_2\tau_8 +
8473652491680\tau_3^2
   + 3265161397760\tau_3\tau_4 - 172696834960\tau_3\tau_5 -
30187385280\tau_3\tau_6
   - 215481388\tau_3\tau_7 - 11665644\tau_3\tau_8 +
314533237920\tau_4^2 - 33235633600\tau_4\tau_5
   - 5774854860\tau_4\tau_6 - 43855192\tau_4\tau_7 -
2157396\tau_4\tau_8 + 875872320\tau_5^2
   + 301721420\tau_5\tau_6 + 2561630\tau_5\tau_7 +
105138\tau_5\tau_8 + 24818780\tau_6^2
   + 673116\tau_6\tau_7 + 10520\tau_6\tau_8 - 10380\tau_7^2 +
924\tau_7\tau_8 - 30\tau_8^2
   - 34610422943232\tau_1^3 - 23549560033152\tau_1^2\tau_2 -
4256637854112\tau_1^2\tau_3
   - 733293440148\tau_1^2\tau_4 + 42798878240\tau_1^2\tau_5 +
4813830704\tau_1^2\tau_6
   + 28229872\tau_1^2\tau_7 + 2175696\tau_1^2\tau_8 -
4526027010240\tau_1\tau_2^2
   - 1627008341488\tau_1\tau_2\tau_3 -
265972385594\tau_1\tau_2\tau_4 + 14544618640\tau_1\tau_2\tau_5
   + 1473032168\tau_1\tau_2\tau_6 + 7097216\tau_1\tau_2\tau_7 +
622688\tau_1\tau_2\tau_8
   - 104213058624\tau_1\tau_3^2 - 31068242652\tau_1\tau_3\tau_4 +
1894244400\tau_1\tau_3\tau_5
   + 30946016\tau_1\tau_3\tau_6 + 847312\tau_1\tau_3\tau_7 +
10416\tau_1\tau_3\tau_8 - 2077171274\tau_1\tau_4^2
   + 273304962\tau_1\tau_4\tau_5 - 4643824\tau_1\tau_4\tau_6 -
6504\tau_1\tau_4\tau_7 + 1448\tau_1\tau_4\tau_8
   - 8972624\tau_1\tau_5^2 - 540360\tau_1\tau_5\tau_6 +
112\tau_1\tau_5\tau_7 - 560\tau_1\tau_5\tau_8 +
109298\tau_1\tau_6^2
   + 8772\tau_1\tau_6\tau_7 - 128\tau_1\tau_6\tau_8 -
88\tau_1\tau_7^2 - 174430354656\tau_2^3
   - 111771006480\tau_2^2\tau_3 - 15318006638\tau_2^2\tau_4 +
525727312\tau_2^2\tau_5
   + 41787228\tau_2^2\tau_6 - 689850\tau_2^2\tau_7 +
12018\tau_2^2\tau_8 - 22407048208\tau_2\tau_3^2
   - 6594418576\tau_2\tau_3\tau_4 + 233297248\tau_2\tau_3\tau_5 +
13952552\tau_2\tau_3\tau_6
   - 175256\tau_2\tau_3\tau_7 + 3320\tau_2\tau_3\tau_8 -
424127412\tau_2\tau_4^2 + 24045634\tau_2\tau_4\tau_5
   + 381734\tau_2\tau_4\tau_6 - 46514\tau_2\tau_4\tau_7 +
22\tau_2\tau_4\tau_8 - 198192\tau_2\tau_5^2 +
93000\tau_2\tau_5\tau_6
   + 4568\tau_2\tau_5\tau_7 - 24\tau_2\tau_5\tau_8 +
16610\tau_2\tau_6^2 + 810\tau_2\tau_6\tau_7 + 4\tau_2\tau_7^2
   - 1345258656\tau_3^3 - 692660864\tau_3^2\tau_4 +
26522288\tau_3^2\tau_5 + 1148744\tau_3^2\tau_6
   - 3504\tau_3^2\tau_7 + 288\tau_3^2\tau_8 -
113020822\tau_3\tau_4^2 + 8373228\tau_3\tau_4\tau_5 +
463430\tau_3\tau_4\tau_6
   + 3118\tau_3\tau_4\tau_7 + 36\tau_3\tau_4\tau_8 -
153344\tau_3\tau_5^2 - 8208\tau_3\tau_5\tau_6 -
96\tau_3\tau_5\tau_7
   + 886\tau_3\tau_6^2 + 84\tau_3\tau_6\tau_7 - 5726384\tau_4^3 +
598808\tau_4^2\tau_5 + 42648\tau_4^2\tau_6
   + 444\tau_4^2\tau_7 - 20576\tau_4\tau_5^2 -
2066\tau_4\tau_5\tau_6 - 16\tau_4\tau_5\tau_7 - 66\tau_4\tau_6^2 +
2\tau_4\tau_6\tau_7
   + 256\tau_5^3 + 32\tau_5^2\tau_6 + 4\tau_5\tau_6^2 +
563452167168\tau_1^4 + 404246383488\tau_1^3\tau_2
   + 45214766208\tau_1^3\tau_3 + 7163567248\tau_1^3\tau_4 -
411309568\tau_1^3\tau_5 - 10337024\tau_1^3\tau_6
   - 131136\tau_1^3\tau_7 - 8960\tau_1^3\tau_8 +
88136227968\tau_1^2\tau_2^2 + 21708131968\tau_1^2\tau_2\tau_3
   + 3187249744\tau_1^2\tau_2\tau_4 -
158966592\tau_1^2\tau_2\tau_5
- 3758528\tau_1^2\tau_2\tau_6
   - 120512\tau_1^2\tau_2\tau_7 - 768\tau_1^2\tau_2\tau_8 +
896258304\tau_1^2\tau_3^2 + 290621136\tau_1^2\tau_3\tau_4
   - 12497280\tau_1^2\tau_3\tau_5 + 267232\tau_1^2\tau_3\tau_6 +
5312\tau_1^2\tau_3\tau_7 + 26342120\tau_1^2\tau_4^2
   - 2086992\tau_1^2\tau_4\tau_5 - 41364\tau_1^2\tau_4\tau_6 -
1216\tau_1^2\tau_4\tau_7 + 44416\tau_1^2\tau_5^2
   - 6464\tau_1^2\tau_5\tau_6 - 1120\tau_1^2\tau_6^2 +
5474101536\tau_1\tau_2^3 + 2328170000\tau_1\tau_2^2\tau_3
   + 259314362\tau_1\tau_2^2\tau_4 - 8140272\tau_1\tau_2^2\tau_5 +
183144\tau_1\tau_2^2\tau_6 - 1536\tau_1\tau_2^2\tau_7
   + 96\tau_1\tau_2^2\tau_8 + 250531648\tau_1\tau_2\tau_3^2 +
59254008\tau_1\tau_2\tau_3\tau_4 - 2310528\tau_1\tau_2\tau_3\tau_5
   + 11792\tau_1\tau_2\tau_3\tau_6 - 1216\tau_1\tau_2\tau_3\tau_7
+
3869472\tau_1\tau_2\tau_4^2 - 269240\tau_1\tau_2\tau_4\tau_5
   - 10286\tau_1\tau_2\tau_4\tau_6 - 320\tau_1\tau_2\tau_4\tau_7 +
2368\tau_1\tau_2\tau_5^2 - 480\tau_1\tau_2\tau_5\tau_6 -
24\tau_1\tau_2\tau_6^2
   + 2764800\tau_1\tau_3^3 - 308064\tau_1\tau_3^2\tau_4 -
11904\tau_1\tau_3^2\tau_5 + 1056\tau_1\tau_3^2\tau_6
   - 192042\tau_1\tau_3\tau_4^2 + 4736\tau_1\tau_3\tau_4\tau_5 +
860\tau_1\tau_3\tau_4\tau_6 - 10498\tau_1\tau_4^3 +
288\tau_1\tau_4^2\tau_5
   + 70\tau_1\tau_4^2\tau_6 + 100182720\tau_2^4 +
57277056\tau_2^3\tau_3 + 3401914\tau_2^3\tau_4 +
6800\tau_2^3\tau_5
   + 5760\tau_2^3\tau_6 - 128\tau_2^3\tau_7 +
11173104\tau_2^2\tau_3^2 + 1326868\tau_2^2\tau_3\tau_4 +
1728\tau_2^2\tau_3\tau_5
   + 2784\tau_2^2\tau_3\tau_6 + 32\tau_2^2\tau_3\tau_7 +
21654\tau_2^2\tau_4^2 + 1800\tau_2^2\tau_4\tau_5 -
64\tau_2^2\tau_5^2
   + 841920\tau_2\tau_3^3 + 180976\tau_2\tau_3^2\tau_4 -
2112\tau_2\tau_3^2\tau_5 - 96\tau_2\tau_3^2\tau_6 +
7748\tau_2\tau_3\tau_4^2
   + 80\tau_2\tau_3\tau_4\tau_5 - 16\tau_2\tau_3\tau_4\tau_6 +
24\tau_2\tau_4^3 + 4\tau_2\tau_4^2\tau_5 + 10368\tau_3^4 +
2592\tau_3^3\tau_4
   + 216\tau_3^2\tau_4^2 + 6\tau_3\tau_4^3 - 4716375552\tau_1^5 -
3618207744\tau_1^4\tau_2 - 200024064\tau_1^4\tau_3
   - 28903104\tau_1^4\tau_4 + 1549312\tau_1^4\tau_5 -
139264\tau_1^4\tau_6 - 837549312\tau_1^3\tau_2^2
   - 106655744\tau_1^3\tau_2\tau_3 - 13843200\tau_1^3\tau_2\tau_4
+ 677888\tau_1^3\tau_2\tau_5 - 16896\tau_1^3\tau_2\tau_6
   - 525312\tau_1^3\tau_3^2 + 197824\tau_1^3\tau_3\tau_4 +
41496\tau_1^3\tau_4^2 - 57285312\tau_1^2\tau_2^3
   - 13249280\tau_1^2\tau_2^2\tau_3 -
1295424\tau_1^2\tau_2^2\tau_4
+ 35072\tau_1^2\tau_2^2\tau_5 - 384\tau_1^2\tau_2^2\tau_6
   - 270592\tau_1^2\tau_2\tau_3^2 - 4608\tau_1^2\tau_2\tau_3\tau_4
+
4492\tau_1^2\tau_2\tau_4^2 - 948384\tau_1\tau_2^4
   - 274560\tau_1\tau_2^3\tau_3 - 17440\tau_1\tau_2^3\tau_4 +
128\tau_1\tau_2^3\tau_5 - 23808\tau_1\tau_2^2\tau_3^2
   - 3680\tau_1\tau_2^2\tau_3\tau_4 - 24\tau_1\tau_2^2\tau_4^2 +
1024\tau_2^4\tau_3 + 256\tau_2^3\tau_3^2 + 17031168\tau_1^6
   + 13400064\tau_1^5\tau_2 + 2881536\tau_1^4\tau_2^2 +
150528\tau_1^3\tau_2^3 + 1536\tau_1^2\tau_2^4\,,
$

\normalsize

\newpage
\section*{Appendix B}

\begin{center}
\begin{longtable}{|c|c|c|c|c|l|r|}
\caption*{{\bf Table.} List of the orbits $\Om_\mathbf{n}$ appearing in \eqref{C_aux}
and orbit decompositions needed to find the coefficients $c_a$}
\label{Table_A} \\
\hline
$a$&$l$&$k$&$\mathbf{n}$&$(\mathbf{n}\cdot\mathbf{n})$&{$M_\mathbf{n}(\tau)$}&$\mu_\mathbf{n}$\\
\endfirsthead

\multicolumn{7}{c}%
{{-- continued from previous page}} \\
\hline
$a$&$l$&$k$&$\mathbf{n}$&$v^2$&$M_\mathbf{n}(\tau)$&$\mu$\\
\hline
\endhead

\hline
\multicolumn{7}{c}%
{{
-- continued on next page}} \\%

\endfoot

\hline
\endlastfoot

\hline
\hline
1&2&1&$[0,0,0,0,0,0,0,0]$&0&1&240\\
\hline
\hline
2&2&1&$[1,0,0,0,0,0,0,0]$&2&$\tau_{1}$&126\\
\hline
\hline
3&2&1&$[0,1,0,0,0,0,0,0]$&4&$\tau_{2}$&84\\
 &3&1,2&$[1,0,0,0,0,0,0,0]$&2&$\tau_{1}$&$2\times 56$\\
\hline
\hline
4&2&1&$[0,0,1,0,0,0,0,0]$&6&$\tau_{3}$&72\\
 &3&1,2&$[0,1,0,0,0,0,0,0]$&4&$\tau_{2}$&$2\times 64$\\
\hline
\hline
5&2&1&$[1,1,0,0,0,0,0,0]$&10&$-126\,\tau_{1}-64\,\tau_{2}
-27\,\tau_{3}-8\,\tau_{4}+\tau_{1}\tau_{2}$&60\\
 &3&1,2&$[0,0,0,1,0,0,0,0]$&8&$\tau_{4}$&$2\times 56$\\
 &4&1,3&$[0,0,0,0,0,0,1,0]$&6&$\tau_{3}$&$2\times 27$\\
 &4&2&$[0,1,0,0,0,0,0,0]$&4&$\tau_{2}$&84\\
\hline
\hline
6&2&1&$[0,0,0,0,1,0,0,0]$&12&$\tau_{5}$&40\\
 &3&1,2&$[1,1,0,0,0,0,0,0]$&10&$-126\,\tau_{1}-64\,\tau_{2}
-27\,\tau_{3}-8\,\tau_{4}+\tau_{1}\tau_{2}$&$2\times 32$\\
 &4&1,3&$[0,0,0,1,0,0,0,0]$&8&$\tau_{4}$&$2\times 28$\\
 &4&2&$[0,0,1,0,0,0,0,0]$&6&$\tau_{3}$&72\\
\hline
\hline
7&2&1&$[0,1,1,0,0,0,0,0]$&18&$-362880-141372\,\tau_{1}-48084\,\tau_{2}-13644\,\tau_{3}$
&40\\
&&&&&$-2668\,\tau_{4}+145\,\tau_{5}+28\,\tau_{6}$&\\
&&&&&$+1512\,\tau_{1}^{2}+456\,\tau_{1}\tau_{2}-27\,\tau_{1}\tau_{3}
-7\,\tau_{1}\tau_{4}+\tau_{2}\tau_{3}$&\\
 &3&1,2&$[1,0,0,1,0,0,0,0]$&16&$4032\,\tau_{1}+1984\,\tau_{2}+792\,\tau_{3}+200\,\tau_{4}
 -16\,\tau_{5}$
 &$2\times 35$\\
&&&&&$-7\,\tau_{6}+\tau_{1}\tau_{4}-32\,\tau_{1}\tau_{2}$&\\
 &4&1,3&$[0,0,0,0,0,1,0,0]$&14&$\tau_{6}$&$2\times 35$\\
 &4&2&$[0,0,0,0,1,0,0,0]$&12&$\tau_{5}$&40\\
 &5&1,4&$[0,0,0,0,1,0,0,0]$&12&$\tau_{5}$&$2\times 16$\\
 &5&2,3&$[0,0,0,1,0,0,0,0]$&8&$\tau_{4}$&$2\times 56$\\
\hline
\hline
8&2&1&$[0,1,0,0,1,0,0,0]$&28&$38707200+13809600\,\tau_{1}+4643920\,\tau_{2}
$&24\\
&&&&&$+1272600\,\tau_{3}+238400\,\tau_{4}-12050\,\tau_{5}-1575\,\tau_{6}-60\,\tau_{7}
$&\\
&&&&&$-283248\,\tau_{1}^{2}-88888\,\tau_{1}\tau_{2}-8064\,\tau_{1}\tau_{3}-1457\,\tau_{1}\tau_{4}
$&\\
&&&&&$+118\,\tau_{1}\tau_{5}+21\,\tau_{1}\tau_{6}
-2156\,\tau_{2}^{2}-656\,\tau_{2}\tau_{3}-56\,\tau_{2}\tau_{4}
$&\\
&&&&&$+\tau_{2}\tau_{5}+1512\,\tau_{1}^{3}+456\,\tau_{1}^{2}\tau_{2}
-27\,\tau_{3}^{2}-6\,\tau_{3}\tau_{4}
$&\\
 &3&1,2&$[0,0,1,1,0,0,0,0]$&26&$13685760+5046624\,\tau_{1}+1663728\,\tau_{2}+459144\,\tau_{3}
$&$2\times 20$\\
&&&&&$+86768\,\tau_{4}-4384\,\tau_{5}-607\,\tau_{6}-21\,\tau_{7}-48384\,\tau_{1}^{2}
 $&\\
&&&&&$-12816\,\tau_{1}\tau_{2}+2808\,\tau_{1}\tau_{3}+697\,\tau_{1}\tau_{4}-16\,\tau_{1}\tau_{5}
$&\\
&&&&&$-6\,\tau_{1}\tau_{6}-512\,\tau_{2}^{2}-88\,\tau_{2}\tau_{3}+21\,\tau_{2}\tau_{4}
+\tau_{3}\tau_{4}
 -32\,\tau_{1}^{2}\tau_{2}$&\\
 &4&1,3&$[1,0,0,0,0,1,0,0]$&24&$-18247680-6704640\,\tau_{1}-2211984\,\tau_{2}-607104\,\tau_{3}
$&$2\times 15$\\
&&&&&$-114196\,\tau_{4}+5784\,\tau_{5}+800\,\tau_{6}+25\,\tau_{7}
$&\\
&&&&&$+72576\,\tau_{1}^{2}+19584\,\tau_{1}\tau_{2}-2088\,\tau_{1}\tau_{3}-448\,\tau_{1}\tau_{4}
$&\\
&&&&&$+\tau_{1}\tau_{6}+384\,\tau_{2}^{2}+96\,\tau_{2}\tau_{3}-7\,\tau_{2}\tau_{4}
$&\\
 &4&2&$[1,0,0,0,1,0,0,0]$&22&$3265920+1212624\,\tau_{1}+403380\,\tau_{2}+111087\,\tau_{3}
$&40\\
&&&&&$+21068\,\tau_{4}-1081\,\tau_{5}-168\,\tau_{6}-4\,\tau_{7}-13608\,\tau_{1}^{2}
 $&\\
&&&&&$-3624\,\tau_{1}\tau_{2}+243\,\tau_{1}\tau_{3}+49\,\tau_{1}\tau_{4}
 +\tau_{1}\tau_{5}-10\,\tau_{2}\tau_{3}
 $&\\
 &5&1,4&$[0,1,0,1,0,0,0,0]$&22&$3179520+1173312\,\tau_{1}+387952\,\tau_{2}+106776\,\tau_{3}
$&$2\times 21$\\
&&&&&$+20144\,\tau_{4}-1024\,\tau_{5}-145\,\tau_{6}-5\,\tau_{7}-12672\,\tau_{1}^{2}$&\\
&&&&&$-3456\,\tau_{1}\tau_{2}+360\,\tau_{1}\tau_{3}+77\,\tau_{1}\tau_{4}
-64\,\tau_{2}^{2}-16\,\tau_{2}\tau_{3}
 $&\\
 &&&&&$+\tau_{2}\tau_{4}$&\\
 &5&2,3&$[0,1,1,0,0,0,0,0]$&18&$-362880-141372\,\tau_{1}-48084\,\tau_{2}-13644\,\tau_{3}$
 &$2\times 16$\\
&&&&&$-2668\,\tau_{4}+145\,\tau_{5}+28\,\tau_{6}$&\\
&&&&&$+1512\,\tau_{1}^{2}+456\,\tau_{1}\tau_{2}-27\,\tau_{1}\tau_{3}
-7\,\tau_{1}\tau_{4}+\tau_{2}\tau_{3}$&\\
 &6&1,5&$[0,0,0,0,0,0,1,0]$&20&$\tau_{7}$&$2\times 10$\\
 &6&2,4&$[0,0,0,0,0,1,0,0]$&14&$\tau_{6}$&$2\times 35$\\
 &6&3&$[0,0,0,0,1,0,0,0]$&12&$\tau_{5}$&40
\end{longtable}
\end{center}

\section*{Appendix C}

The list of the eight eigenfunctions $\varphi_{\mathbf{n}}$ of the Hamiltonian (\ref{h_E8}) which degenerate to variables $\tau$'s in the limit $\nu=0$, thus
becoming the eigenfunctions of the Laplacian written in $\tau-$variables  (\ref{nu=0}),

$ \varphi_{{[1,0,0,0,0,0,0,0]}}\ =\ \tau_1+\,{\frac {240\nu}{29\,\nu+1}}\ ,
\\[5pt]
\; -\ep_{ \left[  1,0,0,0,0,0,0,0 \right]} =2(1+29\,\nu)\,,$\\
\indent

$ \varphi_{{[0,1,0,0,0,0,0,0]}}\ =\ \tau_2+
{\frac{126\nu}{17\,\nu+1}}\tau_1+
{\frac {{15120\nu}^{2}}{(17\,\nu+1)(23\,\nu+1) }}\ , \\[5pt]
-\epsilon_{ \left[  0,1,0,0,0,0,0,0 \right]} = 4(1+23\,\nu)\,,
$
\\ \indent

$ \varphi_{{[0,0,1,0,0,0,0,0]}}\,=\,{\tau_3}+{\frac {84\nu\,(74\,\nu+1)}{(11\,\nu+1)(14\,\nu+1)}}{\tau_1}
+{\frac{84\nu}{11\,\nu+1}}{\tau_2}
+{\frac{6720{\nu}^{2}(74\,\nu+1)}{(14\,\nu+1)(19\,\nu+1)(11\,\nu+1) }}\ ,
\\[5pt]
\, -\epsilon_{ \left[  0,0,1,0,0,0,0,0 \right]} = 6(1+19\,\nu)\,,
$
\\ \indent

$ \varphi_{{[0,0,0,1,0,0,0,0]}}\ =\ {\tau_4}+
\frac{2016\,{\nu}^{2}(107\,\nu+4)}{(11\,
\nu+1)^{2}(13\,\nu+1)}\tau_1+{\frac{48\nu(85
\,\nu+2)}{(11\,\nu+1)^{2}}}\,{\tau_2}+
{\frac {72\nu}{11\,\nu+1}}\tau_3
+{\frac{120960{\nu}^{3}(107\,\nu+4)}
{(17\,\nu+1)(13\,\nu+1)(11\,\nu+1)^{2}}}\ ,
\\[5pt]
-\ep_{ \left[  0,0,0,1,0,0,0,0 \right]} =8(1+17\,\nu)\,,
$
\\ \indent

$ \varphi_{{[0,0,0,0,1,0,0,0]}}\ =\ \tau_5
-
\frac{3024\nu(630504\,{\nu}^{5}+520542\,{\nu}^{4}+111198\,{\nu}^{3}+
9962\,{\nu}^{2}+389\,\nu+5)}
{(13\,\nu+1)(9\nu+1)^{2}(8\nu+1)(19\,\nu+2) (11\,\nu+1)}{\tau_1}
\\
-
{\frac {36\nu(1741896{\nu}^{4}+1095443{\nu}^{3}+179687\,{\nu}^{2}
+10963\nu+211)}{(13\nu+1)(9\nu+1)^2 (8\nu+1)(19\nu+2)}}{\tau_2}
-
{\frac{72\nu(12636{\nu}^{3}+6075{\nu}^{2}+692\nu+22)}
{(13\nu+1)(9\nu+1)^{2}(8\nu+1)}}{\tau_3}
\\
-
{\frac{36\nu(39\nu+11)}{(8\nu+1)(9\nu+1)}}{\tau_4}
+
{\frac {3780{\nu}^{2}}{(9\nu+1)(13\nu+1)}}{\tau_1}^2
+
{\frac {60\nu}{9\nu+1}}{\tau_1}{\tau_2}
-
{\frac{120960{\nu}^{2}(45036{\nu}^{4}+39339{\nu}^{3}+7456{\nu}^{2}
+554\nu+15)}{(13\nu+1)(9\nu+1)^{2}(8\,\nu+1)(19\nu+2)(11\,\nu+1)}}\ ,
\\[5pt]
-\ep_{ \left[  0,0,0,0,1,0,0,0 \right]} = 12(1+14\,\nu)\,,
$
\\ \indent

$ \varphi_{{[0,0,0,0,0,1,0,0]}}\ =\ \tau_6
-
{\frac{12096\nu(987336\,{\nu}^{5}+588549\,{\nu}^{4}+105364\,{\nu}^{3}+
7969\,{\nu}^{2}+259\,\nu+3)}
{(7\,\nu+1)(9\nu+1)(17\,\nu+2)(23\,\nu+3)(8\,\nu+1)(11\,\nu+1)}}{\tau_1}
\\
-
{\frac{288\nu(14578344{\nu}^{5}+8305109{\nu}^{4}+1580835{\nu}^{3}+
131979{\nu}^{2}+4869\nu+64)}
{(7\nu+1)(9\nu+1)(17\nu+2)(23\nu+3)(8\nu+1)(11\,\nu+1)}}{\tau_2}\\
-
{\frac{72\nu(1842104{\nu}^{4}+837215{\nu}^{3}+118203{\nu}^{2}+6317\nu+105)}
{(7\nu+1)(23\,\nu+3)(17\,\nu+2)(8\nu+1)(11\,\nu+1)}}{\tau_3}
-
{\frac{16\nu(8536{\nu}^{2}+2163\nu+65)}
{(7\,\nu+1)(23\nu+3)(8\,\nu+1)}}{\tau_4}
+
{\frac{40\nu}{7\nu+1}}{\tau_5}\\
+
{\frac{2016{\nu}^{2}(32\nu+1)}{(8\nu+1)(7\nu+1)(11\,\nu+1)}}{\tau_1}^{2}
+
{\frac{48\nu(32\nu+1)}{(8\nu+1)(7\nu+1)}}{\tau_1}{\tau_2}
-
{\frac{967680{\nu}^{2}(423144\,{\nu}^{5}+284409\,{\nu}^{4}+57718{\nu}^{3}
+5157\,{\nu}^{2}+209\nu+3)}{(8\nu+1)
(11\,\nu+1)(13\nu+1)(9\nu+1)(17\nu+2)(23\,\nu+3)(7\nu+1)}}\ ,
\\[5pt]
-\ep_{ \left[  0,0,0,0,0,1,0,0 \right]} = 14(1+13\,\nu)\,,
$
\\ \indent

$ \varphi_{{[0,0,0,0,0,0,1,0]}}\ =\ {\tau_7}\\
-
{\frac{60480\nu
(311222520
{\nu}^{7}+356579559 {\nu}^{6}+
182384921{\nu}^{5}+52658812 {\nu}^{4}+8989402 {\nu}^{3}+895571\,{
\nu}^{2}+48135\nu+1080)}{(7\,\nu+1)^{3}(9\nu+1)(19\nu+3)
(13\nu+2)^{2}(8\nu+1)}}{\tau_1}\\
-
{\frac {240\nu(26040982968\,{\nu}^{7}+29565390911\,{\nu}^{6}+
15152846437\,{\nu}^{5}+4379453774\,{\nu}^{4}+748801030\,{\nu}^{3}+
74819267\,{\nu}^{2}+4037053\,\nu+90960 )}{ (
7\nu+1)^{3}(9\nu+1)(19\,\nu+3)(13\nu+2)^{2}(8\,\nu+1)}}{\tau_2}\\
-
{\frac{360\nu(4781918232{\nu}^{7}+5403167819
{\nu}^{6}+2768459713\,{\nu}^{5}+799551260\,{\nu}^{4}+136791946\,{\nu}^
{3}+13703795\,{\nu}^{2}+742429\,\nu+16806 )}{
(7\nu+1)^{3}(9\nu+1)(19\nu+3)(13\nu+2)^{2}(8\nu+1)}}{\tau_3}\\
-
{\frac{80\nu(4045663440{\nu}^{7}+4562526122{\nu}^{6}+2337959690{\nu}^{5}+
675553363{\nu}^{4}+115801043{\nu}^{3}+11643241{\nu}^{2}+633875\nu+14426)
}{(7\nu+1)^{3}(9\nu+1)(19\nu+3)(13\nu+2)^{2}(8\nu+1)}}{\tau_4}\\
+
{\frac{40\nu(8380008\,{\nu}^{5}+
7036517\,{\nu}^{4}+2662161\,{\nu}^{3}+497445\,{\nu}^{2}+44269\,\nu+
1500)}{(7\nu+1)(9\nu+1)(19\nu+3)(13\nu+2)^{2}(8\nu+1)}}{\tau_5}
+
\frac{35\nu(12285{\nu}^{3}+6903\,{\nu}^{2}+1795\,\nu+137)}{(7\nu+1)(9\nu+1)
(19\nu+3)(13\nu+2)} {\tau_6}\\
+
\frac{120960\nu(32144{\nu}^{5}+23315{\nu}^{4}+9330{\nu}^{3}
+1858{\nu}^{2}+172\nu+6)}{(7\nu+1)^{2}(13\nu+2)^{2}(19\nu+3)(8\nu+1)}
{\tau_1}^{2}\\
+
\frac{480\nu(20501208{\nu}^{6}+17126911\,{\nu}^{5}+7265998\,{\nu}^
{4}+1729692\,{\nu}^{3}+227072\,{\nu}^{2}+15305\,\nu+414) }{(7\nu+1)^{2}(9\nu+1)(13\nu+2)^{2}(19\nu+3)(8\nu+1)}
{\tau_1}{\tau_2}
-
\frac{720\nu (56 {\nu}^{2}+37 \nu+3)}{(7\nu+1)(8\nu+1)(13\nu+2)}
{\tau_1}{\tau_3}\\
-
{\frac{35\nu(35\nu+13)}{(7\nu+1)(13\nu+2)}}{\tau_1\tau_4}
+
\frac{1680 {\nu}^{2}}{(7\nu+1)(9\nu+1)}
{\tau_2}^{2}
+
{\frac{40\nu}{7\nu+1}}{\tau_2\tau_3}\\
-
{\frac{2903040\nu (195260688\,{\nu}^{8}+241080658\,{\nu}^{7}+134170588\,{\nu}^{6}+
43246282\,{\nu}^{5}+8595200{\nu}^{4}+1065024\,{\nu}^{3}+80145{\nu}^{2}+
3355\nu+60)}{(13\nu+2)^{2}(8\nu+1)(9\nu+1)(19\nu+3)(11\nu+1)(7\nu+1)^{3}}}
\ ,\\[5pt]
-\epsilon_{ \left[  0,0,0,0,0,0,1,0 \right]} = 20(1+11\,\nu)\,,
$
\\ \indent


$ \varphi_{{[0,0,0,0,0,0,0,1]}}\ =\ {\tau_8}
+\\
\frac{414720\nu}{\left( 2+11\,\nu \right)^{2} \left( 5\,\nu+1 \right)^{4} \left( 7\,\nu+1 \right)^{2} \left( 16
\,\nu+3 \right)  \left( 21\,\nu+4 \right)  \left( 17\,\nu+3 \right)
\left( 13\,\nu+2 \right)}
{\scriptstyle \left(
31712497866000\,{\nu}^{11} +
68059224432675\,{\nu}^{10} +
\right.}
\\
{\scriptstyle\left.
66833164151565\,{\nu}^{9}  +
40155845946657\,{\nu}^{8} +
16408038101157\,{\nu}^{7} +
4758350333107\,{\nu}^{6} +
991892916403\,{\nu}^{5} +
\right.}
\\
{\scriptstyle\left.
147654773755\,{\nu}^{4} +
15306171631\,{\nu}^{3} +
1048498510\,{\nu}^2 +
42607780\nu +
776760
\right)
-}
\\
\frac{\nu\,{\tau_7}}{\left( 21
\,\nu+4 \right)  \left( 7\,\nu+1 \right)  \left( 5\,\nu+1 \right) ^{4}
 \left( 2+11\,\nu \right)}
{\scriptstyle
\left(
80243625{\nu}^{6} +
104650525\,{\nu}^{5} +
58254520\,{\nu}^{4} +
19039926\,{\nu}^{3} +
3634709\,{\nu}^{2} +
360941\,\nu +
14154
\right)
-
}
\\
\frac{10\nu\,{\tau_6}}{ \left( 21\,\nu+4 \right)
 \left( 16\,\nu+3 \right)  \left( 7\,\nu+1 \right)  \left( 5\,\nu+1
 \right) ^{4} \left( 2+11\,\nu \right) ^{2}}
{\scriptstyle
\left(
36728538000\,{\nu}^{8} +
61474586075\,{\nu}^{7} +
\right.}
\\
{\scriptstyle \left.
45470399905{\nu}^{6} +
20097488322\,{\nu}^{5} +
5737716840\,{\nu}^{4} +
1063013277\,{\nu}^{3} +
122932983{\nu}^{2} +
8025190\nu +
224528
\right)
-
}
\\
\frac{160\nu {\tau_5}}{\left( 17\,\nu+3 \right)  \left( 21\,\nu+4
 \right)  \left( 16\,\nu+3 \right)  \left( 7\,\nu+1 \right)  \left( 5
\,\nu+1 \right) ^{4} \left( 2+11\,\nu \right) ^{2}}
{\scriptstyle
\left(
287586684000\,{\nu}^{9} +
532807847250\,{\nu}^{8} +
442238671740\,{\nu}^{7} +
221304181654\,{\nu}^{6}+
\right.}\\
{\scriptstyle \left.
73267827503\,{\nu}^{5} +
16435840960\,{\nu}^{4} +
2468093966\,{\nu}^{3} +
236963590\,{\nu}^{2} +
13105703\,\nu +
316434 \right)
+
}
\\
\frac{80\nu
{\tau_4}}{\left( 17\,\nu+3
 \right)  \left( 21\,\nu+4 \right)  \left( 16\,\nu+3 \right)
 \left(7\nu+1 \right) \left(5\nu+1 \right)^{4} \left( 2+11\,\nu
 \right) ^{2}}
{\scriptstyle
\left(
11362507002000\,{\nu}^{9} +
21040524174075\,{\nu}^{8} +
17465560810170\,{\nu}^{7} +
\right.}
\\
{\scriptstyle\left.
8741275255117\,{\nu}^{6}  +
2894996036768\,{\nu}^{5} +
649680782473\,{\nu}^{4} +
97592599610\,{\nu}^{3} +
9372446035\,{\nu}^{2} +
518463836\,\nu +
12519516
\right)
+
}
\\
\frac{360\nu{\tau_3}
}{\left( 2+11\,\nu \right) ^{2} \left( 5\,\nu+1 \right)^{4} \left( 7
\,\nu+1 \right) ^{2} \left( 16\,\nu+3 \right)  \left( 21\,\nu+4
 \right)  \left( 17\,\nu+3 \right)  \left( 13\,\nu+2 \right)}
{\scriptstyle
\left(
1217275708110000\,{\nu}^{11} +
2614590200795825\,{\nu}^{10} +
\right.}
\\
{\scriptstyle\left.
2566185152952895\,{\nu}^{9} +
1541285426902427\,{\nu}^{8} +
629439297115555\,{\nu}^{7} +
182398178713341\,{\nu}^{6} +
37987084347873\,{\nu}^{5} +
\right.}
\\
{\scriptstyle\left.
5649252328153\,{\nu}^{4} +
585014696297\,{\nu}^{3} +
40034832310\,{\nu}^{2} +
1625479460\,\nu +
29613864
\right)
+
}
\\
\frac{240\,\nu\,{\tau_2}}{ \left( 13\,\nu+2 \right)
 \left( 17\,\nu+3 \right)  \left( 7\,\nu+1 \right)  \left( 16\,\nu+3
 \right)  \left( 2+11\,\nu \right) ^{2} \left( 5\,\nu+1 \right) ^{4}
 \left( 21\,\nu+4 \right)}
{\scriptstyle
\left(
943289442942000\,{\nu}^{10} +
1890679995233225\,{\nu}^{9} +
\right.}
\\
{\scriptstyle\left.
1718559215503880\,{\nu}^{8} +
949543222464315\,{\nu}^{7} +
352718586334054\,{\nu}^{6} +
91236988028507\,{\nu}^{5} +
16485156918772\,{\nu}^{4} +
\right.}
\\
{\scriptstyle\left.
2038447895205\,{\nu}^{3} +
164128225478\,{\nu}^{2} +
7737426812\,\nu +
161683752
\right)
+
}
\\
\frac{8640\nu{\tau_1}
}
{(11\nu+2)^{2}(5\nu+1)^{4}(7\nu+1)^{2}
(16\nu+3)(21\nu+4)(17\nu+3)(13\nu+2)}
{\scriptstyle
\left(
541228676310000{\nu}^{11} +
1161538631434125{\nu}^{10} +
\right.}
\\
{
\scriptstyle \left.
1140551008849875{\nu}^{9} +
685499100975535{\nu}^{8} +
280292653233119{\nu}^{7} +
81372038745921{\nu}^{6} +
16985484665469{\nu}^{5} +
\right.}
\\
{
\scriptstyle
\left.
2532327847389{\nu}^{4} +
262907743293{\nu}^{3} +
18035930558\,{\nu}^{2} +
733923940\nu +
13396776
\right)
-}
\\
{\scriptstyle
\frac{6\nu(35\nu+19)}{(5\nu+1)^{2}}{\tau_3}\,{\tau_4}
-
\frac{72\nu(145{\nu}^{2}+98\nu+9)}{(7\nu+1)(5\nu+1)^{2}}{\tau_3}^{2}
+
\frac{24\nu}{5\nu+1}{\tau_2}\,{\tau_5}
-
}
\\
\frac{\nu(2276925{\nu}^{4}+1977050{\nu}^{3}+555508{\nu}^{2}+68806\nu+3311)
{(7\nu+1)(5\nu+1)^{3}(21\nu+4)}{\tau_2}\,{\tau_4}}\\
-
\\
\frac{32\nu(768306000{\nu}^{7}+1156535125{\nu}^{6}+717875380{\nu}^{5}
+247730074{\nu}^{4}+51665823{\nu}^{3}+6462602{\nu}^{2}+443993\nu+12803)
}{(7\nu+1)(16\nu+3)(11\nu+2)(5\nu+1)^{4}(21\nu+4)} {\tau_2}\,{\tau_3}
\\
{\scriptstyle -}
\frac{288\nu(300762000\,{\nu}^{7}+450801075
\,{\nu}^{6}+276732990\,{\nu}^{5}+95786616\,{\nu}^{4}+20239042{\nu}^{3}
+2564033{\nu}^{2}+177152\,\nu+5092)}{
(7\nu+1)(16\nu+3)(11\nu+2)(5\nu+1)^{4}(21\nu+4)}{\tau_2}^{2}
\\
{\scriptstyle +}
\frac{4\nu(2975{\nu}^{3}+2170{\nu}^{2}+817\nu+86)}{(7\nu+1)(5\nu+1)^{3}}
{\tau_1}{\tau_6}
+
\frac{8(169575\nu{\nu}^{4}+146030{\nu}^{3}+58468{\nu}^{2}+10030\nu+593)
}{(7\nu+1)(5\nu+1)^{3}(11\nu+2)}{\tau_1}\,{\tau_5}
\\
{\scriptstyle -}
\frac{\nu(27492822000{\nu}^{7}+39665858925{\nu}^{6}+26425044185{\nu}^{5}
+9798575016{\nu}^{4}+2104815742{\nu}^{3}+260135833{\nu}^{2}+17270457\nu
+482642)}{(7\nu+1)(16\nu+3 )(11\nu+2)(5\nu+1)^{4}(21\nu+4)}{\tau_1}\,{\tau_4}
\\
{\scriptstyle -}
\frac{72\nu(2703918000\,{\nu}^{7}+3988109425{\nu}^{6}+2618867645{\nu}^{5}
+960198076{\nu}^{4}+205182062{\nu}^{3}+25247533{\nu}^{2}+1661957\nu
+45702)}
{(7\nu+1)(16\nu+3)(11\nu+2)(5\nu+1)^{4}(21\nu+4)}{\tau_1}\,{\tau_3}
\\
{\scriptstyle -}
\frac{24192\nu{\tau_1}^{2}
}
{(17\nu+3) (21\nu+4)(16\nu+3)(7\nu+1)^{2}
(5\nu+1)^{4}(11\nu+2)^{2}}
{\scriptstyle
\left(
342813408000{\nu}^{10} +
685862951300{\nu}^{9} +
618894401705{\nu}^{8} +
\right.}
\\
{
\scriptstyle \left.
337437179828{\nu}^{7} +
122503429953{\nu}^{6} +
30671435796{\nu}^{5} +
5326124025{\nu}^{4} +
630491818{\nu}^{3} +
48546661{\nu}^{2} +
2190970\nu +
43944
\right)
-
}
\\
\frac{96\nu{\tau_1}{\tau_2}}{(17\nu+3)(21\nu+4)(16\nu+3)(7\nu+1)(5\nu+1)^{4}
(11\nu+2)^{2}}
{\scriptstyle
\left(
4118689806000{\nu}^{9} +
7629472330025{\nu}^{8} +
6304799793280{\nu}^{7} +
\right.}
\\
{
\scriptstyle \left.
3105236849563{\nu}^{6} +
998271056054{\nu}^{5} +
215044644679{\nu}^{4} +
30799297148{\nu}^{3} +
2813638949{\nu}^{2} +
148263454\nu +
3424848
\right)
+
}
\\
\frac{4032\nu(6650\,{\nu}^{4}+5405{\nu}^{3}+2173{\nu}^{2}+340\nu+18)}
{(7\nu+1)(5\nu+1)^{3}(11\nu+2)}{\tau_1}^{3} {\scriptstyle +}
\frac{96\nu(77525{\nu}^{4}+61805{\nu}^{3}+23733{\nu}^{2}+3805\nu+208)}
{(7\nu+1)(5\nu+1)^{3}(11\nu+2)}{\tau_1}^{2}{\tau_2} {\scriptstyle +}
\frac{720{\nu}^{2}}{(5\nu+1)(7\nu+1)}{\tau_1}{\tau_2}^{2}
\ , \\[5pt]
-\epsilon_{ \left[  0,0,0,0,0,0,0,1 \right]} = 30(1+9\,\nu)\,.
$

\end{document}